\documentclass[structabstract]{aa}  
\usepackage{graphicx}
\usepackage{ gensymb }
\usepackage{ tipa }
\usepackage{ textcomp }
\usepackage{ wasysym }
\usepackage{float}

\usepackage{rotating}
\usepackage{txfonts}
\usepackage{lscape}
\begin{document}
   \title{Diagnostic value of far-IR water ice features in T Tauri disks}

 \author{I. Kamp
	\inst{1}
	\and
	A. Scheepstra\inst{1}
	\and
  	M. Min\inst{2}
	\and
	L. Klarmann\inst{2}
        \and
        P. Riviere-Marichalar \inst{3}
	}

                 \institute{Kapteyn Astronomical Institute, University of Groningen, The Netherlands\\
\email{kamp@astro.rug.nl}
     		\and
                 Anton Pannekoek Institute, University of Amsterdam, The Netherlands 
                \and
                Department of Surfaces, Coatings, and Molecular Astrophysics, Spanish National Research Council, Madrid, Spain            		
       			}
 

 
  \abstract
 {}
  {This paper investigates how the far-IR water ice features can be used to infer properties of disks around T Tauri stars and the water ice thermal history. We explore the power of future observations with SOFIA/HIRMES and SPICA's proposed far-IR instrument SAFARI.}
   {A series of detailed radiative transfer disk models around a representative T Tauri star are used to investigate how the far-IR water ice features at $45$ and $63~\mu$m change with key disk properties: disk size, grain sizes, disk dust mass, dust settling, and ice thickness. In addition, a series of models is devised to calculate the water ice emission features from warmup, direct deposit and cooldown scenarios of the water ice in disks. }
   {Photodesorption from icy grains in disk surfaces weakens the mid-IR water ice features by factors 4-5. The far-IR water ice emission features originate from small grains at the surface snow line in disks at distance of 10-100~au. Unless this reservoir is missing in disks (e.g. transitional disks with large cavities), the feature strength is not changing. Grains larger than 10~$\mu$m do not contribute to the features. Grain settling (using turbulent description) is affecting the strength of the ice features by at most 15\%. The strength of the ice feature scales with the disk dust mass and water ice fraction on the grains, but saturates for dust masses larger than $10^{-4}$~M$_\odot$ and for ice mantles that increase the dust mass by more than 50\%. The various thermal histories of water ice leave an imprint on the shape of the features (crystalline/amorphous) as well as on the peak strength and position of the $45~\mu$m feature. SOFIA/HIRMES can only detect crystalline ice features much stronger than simulated in our standard T Tauri disk model in deep exposures (1 hr). SPICA/SAFARI can detect the typical ice features in our standard T Tauri disk model in short exposures (10 min).}
   {The sensitivity of SPICA/SAFARI will allow the detailed study of the 45 and 63~$\mu$m water ice feature in unbiased surveys of T Tauri stars in nearby star forming regions and an estimate of the mass of their ice reservoir. The water ice emission features carry an imprint of the thermal history of the ice and thus can distinguish between various formation and transport scenarios. Amorphous ice at 45~$\mu$m that has a much broader and flatter peak could be detected in deep surveys if the underlying continuum can be well characterized and the baseline stability of SAFARI is better than a few percent.}

   \maketitle
%

\section{Introduction}

During planet formation, ice plays a key role in multiple ways. Water ice enhances the solid mass reservoir available for planet formation, it improves the sticking properties of small dust grains for coagulation into larger bodies \citep[e.g.][]{Wada2007, Wettlaufer2010, Wada2013, Gundlach2015}. More indirectly, ices are an extra opacity source thus affecting the disk thermal structure. In addition, they could play an important role in providing the reservoir to make larger organic molecules \citep[see review by][]{MunozCaro2013}. 

In Herbig disks, water ice was first detected by \citet{Malfait1999} in the ISO/SWS and LWS spectra of HD\,142527, a detection followed up later by Herschel/PACS \citep{Min2016b}. The disk contains an outer reservoir of highly crystalline ice (beyond $\sim\!130$~au), which comprises a major fraction of the available oxygen (\mbox{$\sim\!80$\%}). \citet{Chiang2001} report the detection of the 45~$\mu$m water ice feature in ISO/LWS spectra of CQ~Tau and AA~Tau; the spectra unfortunately lack full wavelength coverage of this feature. Later Herschel/PACS spectra of a small sample of T Tauri stars also suffer from partial coverage of the ice features and baseline stability issues. However, \citet{McClure2012} were able to report a detection of the 63~$\mu$m crystalline ice feature in GQ~Lup and \citet{McClure2015} in Haro~6-13 and DO~Tau.

The history and evolution of the ices in disks remains unclear. \citet{Visser2009} show from cloud collapse simulations that water ice is incorporated into the outer reservoir of a protoplanetary disk without desorption (never heated above the sublimation temperature). Recent observations of the comet 67P/Churyumov-Gerasimenko with the Rosetta mission show high abundances of the noble gas argon suggesting that the ice of that comet has never been heated above $\sim\!20$~K \citep{Balsiger2015}. On the other hand, the high fraction of cold crystalline ice seen in the disk around HD\,142527 requires some heating and/or mixing processes within the disk.

Photodesorption processes can remove ices from grains in the surface layers of cold outer disks \citep{Dominik2005, Woitke2009b}, thereby creating a cold water vapour reservoir. This cold water vapour was subsequently detected by Herschel/HIFI in the disks around TW~Hya and DG~Tau \citep{Hogerheijde2011, Podio2013}. After establishing the relevance of non-thermal desorption, \citet{Min2016b} model for the first time the water ice reservoir self consistently in the disk around a Herbig star with the 2D thermal disk structure and photodesorption. They use an ice line description that is calibrated on  thermo-chemical disk models.

In this work, we expand the disk modeling to T Tauri disks. We use the reference T Tauri disk model presented in \citet{Woitke2016} (Sect.~\ref{sec:models}). By varying key disk parameters such as the disk size, dust mass, grain sizes, dust settling, and ice thickness, we exploit the diagnostic value of the far-IR water ice features at 45 and 63~$\mu$m (Sect.~\ref{sec:results}). In addition, we explore whether we can observationally distinguish the three scenarios of warm-up, in-situ formation, and cool-down for water ice. In Sect.~\ref{sec:future}, we discuss our results in the context of the approved SOFIA/HIRMES instrument and the proposed SPICA/SAFARI instrument.

\section{The disk models}
\label{sec:models}

\begin{table}[th]
\caption{Basic model parameters for the reference T Tauri disk.}
\begin{tabular}{lll}
\hline
\hline\\[-3mm]
 Quantity       & Symbol & Value \\
\hline\\[-3mm]
Stellar mass & $M_\ast$ & 0.7~M$_\odot$\\
Effective temperature & $T_{\rm eff}$ & 4000~K\\
Stellar luminosity & $L_\ast$ & 1.0~L$_\odot$\\
FUV excess & $f_{\rm UV}$ & 0.01\\
                       & $p_{\rm UV}$ & 1.3\\
\hline\\[-3mm]
Disk dust mass$^1$ & $M_{\rm dust}$ & $3.3\,10^{-4}$~M$_\odot$\\
Inner disk radius & $R_{\rm in}$ & 0.07~au\\
Outer disk radius$^2$ & $R_{\rm out}$ & 700~au\\
Tapered edge radius & $R_{\rm taper}$ & 100~au\\
Column density power index & $\epsilon$ & 1.0\\
Reference radius & $R_0$ & 100~au \\
Scale height at $R_{0}$ & $H_0$ & 10.0~au \\
Disk flaring power index & $\beta$ & 1.15 \\
\hline\\[-3mm]
Minimum dust particle radius & $a_{\rm min}$ & 0.05 $\mu$m\\
Maximum dust particle radius & $a_{\rm max}$ & 3000.0 $\mu$m \\
Dust size dist.\ power index & $a_{\rm pow}$ & 3.5\\
Turbulent mixing parameter & $\alpha_{\rm settle}$ & $10^{-2}$\\
Max.\ hollow volume ratio & $V_{\rm hollow}^{\rm max}$ & 80\% \\
Dust composition & Mg$_{0.7}$Fe$_{0.3}$SiO$_3$\hspace*{-2mm} & 60\% \\
(volume fractions) & amorph.\ carbon\hspace*{-2mm}  & 15\% \\
                            & porosity & 25\% \\
Grain material density & $\rho_{\rm grain}$ & 2.076~g~cm$^{-3}$ \\
\hline
\end{tabular}
\label{tab:diskparameters}
\tablefoot{(1) The disk mass is a factor 3.3 higher than in the original \citet{Woitke2016} model. (2) The outer radius is defined as the radius where the surface density column drops to $N_{\rm \langle H\rangle, ver}=10^{20}$~cm$^{-2}$.} 
\end{table}

All disk models are calculated using the 3D Monte Carlo radiative transfer code MCMax \citep{Min2009}. The code uses a grain size distribution where the fraction of ices is a free parameter. We use the DIANA\footnote{DIANA is the EU FP7 project ``Disc Analysis'' (PI: P.\ Woitke) that developed tools for the interpretation of multi-wavelengths observations of protoplanetary disks and applied them to a large number of protoplanetary disks using a consistent approach.} grain opacities \citep{Min2016a} which were constructed to fit simultaneously observed thermal and scattered light properties in protoplanetary disks. The ice opacities are taken from \citep{Smith1994} and for amorphous ice from \citet{Li1998}; the code can use either opacities at a fixed temperature or temperature dependent opacities. In the following we use 2D disk models and solve the continuum radiative transfer iteratively together with the location of the water ice reservoir. For the latter, we use the formalism developed in \citet{Min2016b}, where the snow line depends on the temperature, pressure and UV radiation field (thermal and non-thermal desorption processes); the oxygen abundance is constrained by setting the total ice fraction in the disk $f_{\rm ice}$. Typically, three iterations are needed for the disk structure and spectral energy distribution (SED) to converge.

Table~\ref{tab:diskparameters} summarizes the key parameters of the reference T Tauri disk model from \citet{Woitke2016}. We focus here on passive irradiated disks where viscous heating plays a negligible role in the midplane close to the star. Since we investigate in this work the far-IR water ice features typically originating from beyond 10~au, the precise radial position of the midplane snowline is not expected to affect our results.

The fraction of ice is fixed to 0.8 for the reference model and the disk parameter series including the thermal history of ice series. However, we include one model series where we explore the impact of the water ice fraction on the SED and the strength of the water ice features. The total dust mass is spread over silicates, carbon and vacuum (60, 15 and 25\% respectively). The fraction of ice is defined as 
\begin{equation}
M_{\rm ice}=f_{\rm ice} M_{\rm dust}\,\,\, .
\end{equation}
In this way, changing the ice fraction does not change the underlying bare grain opacities. Having a non-zero ice fraction adds a mantle around each grain and the total mass of extra ice is spread using the underlying bare grain size distribution. The values of the first, second and third moment of the bare grain size distribution are $\langle a \rangle \!=\! 8.33\,10^{-2}~{\rm \mu m}$, $\langle a^2 \rangle \!= \! 1.245\,10^{-2}~{\rm \mu m^2}$, $\langle a^3 \rangle \!= \! 1.525\,10^{-1}~{\rm \mu m^3}$. The ice fractions chosen here (see Table~\ref{tab:variedparameters}) explore a reasonable range since thermo-chemical disk models suggest that the grains grow almost a factor two in size due to the ice mantles \citep{Chaparro2016}. The fractions of 0.4, 0.8, 1.2, 1.6 and 2.0 correspond to a grain size increase of a factor 1.9, 2.2, 2.4, 2.5 and 2.6 respectively assuming an average water ice density of 1~g~cm$^{-3}$. 

\subsection{Model series}

In order to isolate disk effects on the SED and the 45 and $63~\mu$m water ice features, the base choice for opacities is crystalline ice at 140~K unless stated otherwise. For comparison, the reference disk model is also calculated using amorphous ice opacities and without taking photodesorption into account. For all model series, only one of the parameters of the disk is altered at a time. This allows us again to isolate the effect of several properties of the disk on the water ice features. Table~\ref{tab:variedparameters} shows all the parameters that were varied; the values of the reference disk model are found in Table~\ref{tab:diskparameters}. 

\begin{table}[th]
	\caption{Varied parameters for the disk models.}
	\begin{tabular}{lll}
		\hline
		\hline\\[-3mm]
		Quantity       & Symbol & Values \\
		\hline\\[-3mm]
		Inner radius & $R_{\rm in}$ & 0.07, 0.2, 1, 10, 30, 50 au\\
		Tapering-off radius & $R_{\rm taper}$ & 50, 100, 200 au\\
               Disk dust mass & $M_{\rm dust}$ & $10^{-3}$, $5\,10^{-4}$, $3.3\,10^{-4}$, \\
                                        &                          & $2\,10^{-4}$, $10^{-4}$, $10^{-5}$, $10^{-6}$, \\
                                        &                         & $10^{-7}$~M$_\odot$ \\
		Minimum grain size & $a_{\rm min}$ & 0.05, 0.5, 2, 30, 50 $\mu$m\\
		Turbulent mixing & $\alpha_{\rm settle}$ & $10^{-2}$, $10^{-3}$, $10^{-4}$\\
		fraction water ice & $f_{\rm ice}$ & 0.4, 0.8, 1.2, 1.6, 2.0\\
		\hline
	\end{tabular}
	\label{tab:variedparameters}
\end{table}

An additional series was done where the composition of the water ice is varied, based on its thermal history. We used the same series that are described by \citet{Smith1994}, which consist of\\[-5mm]
\begin{itemize}
	\item Series 1: Warmup from 10 to 150~K\\[-3mm]
	\item Series 2: Direct deposit between 10 and 150~K\\[-3mm]
	\item Series 3: Cooldown from 140 to 10~K\\[-5mm]
\end{itemize}
The warmup series refers to a scenario where the water ice is formed in cold, optically thick regions. After its formation it moves to warmer areas in the disk. We mimick this by using for each temperature interval in the disk the corresponding ice opacity from the warmup series of \citet{Smith1994}. The water ice in the direct deposit scenario is formed locally and stays in the same thermal region. Also here, we change the ice opacity as a function of disk temperature using this time the direct deposity series from Smith's work. In the cooldown scenario, the water ice is formed in warm areas close to the midplane snowline ($\sim\!0.5$~au, around 140~K), then moves outwards to colder areas in the disk. This is simulated using the same methodology as the other two series, but now with the cooldown opacity series.

For each disk model, we calculate the disk thermal structure, extent of the ice reservoir and spectral energy distribution self-consistently. In the following, we describe the analysis of the strength of the water ice features.

\subsection{Analysis of water ice features}
\label{Sect:Analysis}

The trend that we are seeing in the shape of the water ice features between our different disk models is quantified by using a polynomial best fit to the continuum of the SED. Given the strong changes in SED shape related to disk geometry, the placement of the continuum can be very uncertain. Also, observers have no possibility to choose a more refined method since the underlying continuum could not be estimated independently like we could do in the models. To avoid these problems, we chose the simplest solution, a linear fit between $40.0$ and $57.5~\mu$m that covers the $45~\mu$m water ice feature. The $63~\mu$m feature cannot be analyzed this way because of the broad and weak nature of the feature and the uncertainty of the shape of the underlying continuum. 

\begin{figure}[thb]
	\includegraphics[width=9.5cm]{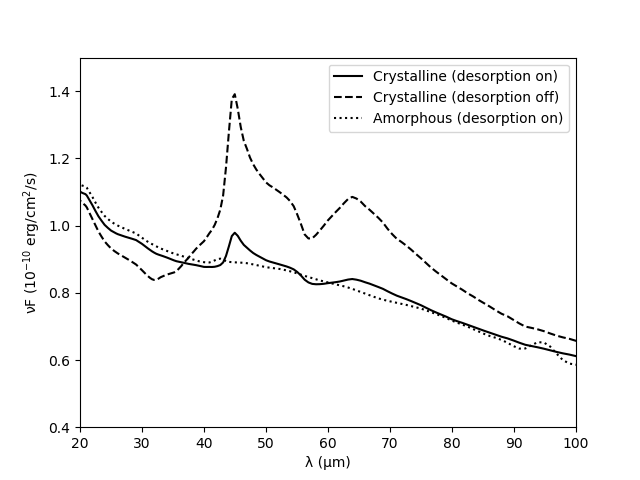} 
	\caption{SED of the reference disk model using crystalline ice opacities (140~K) with photodesorption switched on (solid line) and off (dashed line) and amorphous ice (photodesorption on, dotted line).}
	\label{fig:ZoomedSED_reference-photodesoffl}
\end{figure}

We define the height of the $45~\mu$m water ice features $S_{\rm \lambda,peak}$ as
\begin{equation}\label{fluxSED-fluxContinuum}
\hspace{2.7cm} S_{\rm \lambda,peak} = \frac{F_{\rm \lambda,peak} - F_{\rm \lambda,cont}}{F_{\rm \lambda,cont}}
\end{equation} 
where $F_{\rm \lambda,cont}$ and $F_{\rm \lambda,peak}$ are the continuum and peak flux of the SED respectively. Since the height of the features is sensitive to the definition of the continuum, trends are only considered real if they are significantly larger than the error bars. We estimate the errors, by also using higher degree polynomials to fit the continuum and derive a peak strength. Due to the specific shape of the continuum over the 30-60~$\mu$m wavelength range (mostly concave), the polynomials systematically provide higher peak values (typically $10-25$\%) compared to the linear fit. The uncertainty within the polynomials is smaller, typically only 10\%.

\section{Results}
\label{sec:results}

\subsection{Ice opacities and photodesorption}

Photodesorption is extremely important for the strength of the water ice features. Taking this effect not into account will overestimate their strengths by a factor 4-5 (Fig.~\ref{fig:ZoomedSED_reference-photodesoffl}). Also, amorphous water ice produces features that are a factor $\sim\!7$ weaker than those of crystalline ice, to the extent that the peak strength for the $45~\mu$m feature becomes only $0.02$. Hence, in all subsequent parameter studies except for the `thermal history' series, we adapt crystalline ice opacities (140~K). Note that this does not imply that we think water ice in disks has a temperature of 140~K, but it merely ensure a straightforward and clean analysis of the features. The more realistic `thermal history' water ice scenarios will be discussed in Sect.~\ref{sect:dustice-changefeature}. It is also important to note that the underlying shape of the continuum as well as its absolute value changes depending on the ice opacity choice. This is due to the self-consistent solution of the dust radiative transfer.

\subsection{Location of the water ice reservoir}

We focus in the following on the reference disk model to explain the location of the water ice reservoir and to illustrate where the 45 and 63~$\mu$m ice features originate with respect to the continuum optical depth at those wavelengths.

\begin{figure}[thb]
         \vspace*{-5mm}
	\includegraphics[width=10cm]{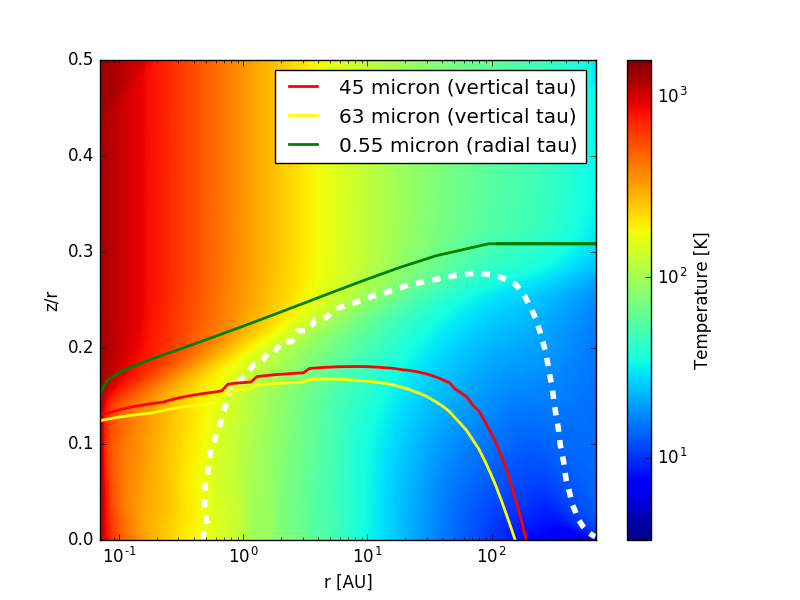} 
	\caption{The 2D temperature disk structure for the reference disk model. The thick white dashed contour outlines the water ice reservoir. The black dashed lines are the 160 and 100~K temperature contours. The thick red and yellow contours denote the vertical optical depth $\tau\!=\!1$ at 45 and 63~$\mu$m. The green and orange contours denote the radial optical depth $\tau\!=\!1$ at $0.55$ and 3~$\mu$m respectively.}
	\label{fig:2Dtemp_referencemodel}
\end{figure}

Fig. \ref{fig:2Dtemp_referencemodel} shows the 2D temperature disk structure of the reference disk model. The thick white, dashed contour outlines the water ice reservoir of the disk (snowline). Note that at the outer radius of the disk, even though these are very cold regions, there is no water ice present. This is because of the interstellar radiation field, which causes photodesorption of ice mantles on dust grains in low density regions. The snowline is defined by the equation derived by \citet{Min2016b}, an equilibrium between the vapor pressure and thermal/non-thermal desorption processes. The black lines show the region where the temperature is between 100 and 160~K. This is roughly the area where the crystallization of water ice is happening \citep{Smith1994}. The red and yellow lines show the location where the photons can escape the disk without further absorption and re-emission. Photons emitted below the vertical optical depth $\tau\!=\!1$ line at 45 and $63~\mu$m will get absorbed again, and observers can therefore not obtain information about the structure of the disk below this region. The green and orange lines show where stellar photons with a wavelength of 0.55 and $3~\mu$m, respectively, first interact with the disk. The stellar emission peaks at roughly $0.55~\mu$m. Stellar photons with a wavelength of $3~\mu$m are not directly absorbed by the ice reservoir, if the region has a hard cut-off and therefore the $3~\mu$m water ice feature may not show in observations of protoplanetary disks with simple standard geometries (full disk, no gaps, holes, no strong settling). 

The 45 and $63~\mu$m water ice features in the SED of protoplanetary disks, originate between the snowline and the vertical optical depth $\tau\!=\!1$ lines at 45 and $63~\mu$m. The following sections will use various model series to understand in more detail the radial range over which the ice feature originate.


\subsection{The peak strength of the $45~\mu$m feature}

Table \ref{tab:heightfeatures} shows the height of the 45~$\mu$m water ice feature $S_{\rm45, peak}$ for the different disk model series. It is important to recall that the peak strength carries an error of typically less than $10$\% (from use of different polynomial methods). 

In the following sections, we discuss in detail the different parameter series in two groups: Parameters that do not change the water ice features and those which do. This enables us to assess the key science that can be learned from studying these features in the future.

\begin{table}[h]
	\caption{Peak strength of the $45~\mu$m water ice feature in the reference model and the various T Tauri disk model series.}
	\begin{tabular}{lll}
		\hline
		\hline\\[-3mm]
		Model series       & Value & $S_{\rm45, peak}$ \\
		\hline\\[-3mm]
		Reference (crystalline) & - &  0.134\\
		Reference (amorphous) & - &  0.020\\
		\hline\\[-3mm]
		$R_{\rm in}$ [au] & 0.07  & 0.133\\
		& 0.2  & 0.138\\
		& 1  & 0.108\\
		& 10 &  * \\
		& 30  & * \\
		& 50  & * \\
		\hline\\[-3mm]
		$R_{\rm taper}$ [au] & 50  &  0.141\\
		& 100  & 0.130\\
		& 200  &  0.128\\
		\hline\\[-3mm]
		$a_{\rm min}$ [$\mu$m] & 0.05 & 0.131 \\
		& 0.5 &  0.132\\
		& 2  &   0.111\\
		& 10  &  0.043\\
		& 30  &  $8\,10^{-4}$\\
		& 50  &  0.0\\
                \hline\\[-3mm]
                $M_{\rm dust}$ [M$_\odot$] & $10^{-3}$ & 0.138\\
                & $5\,10^{-4}$ & 0.141 \\
                & $3.3\,10^{-4}$ & 0.132\\
                & $2\,10^{-4}$ & 0.127\\
                & $10^{-4}$ & 0.127\\
                & $10^{-5}$ & 0.107\\
                & $10^{-6}$ & 0.050\\
                & $10^{-7}$ & 0.0\\
		\hline\\[-3mm]
		$\alpha_{\rm settle}$ & $10^{-2}$ & 0.132 \\
		& $10^{-3}$ & 0.146 \\
		& $10^{-4}$ & 0.157 \\
		\hline\\[-3mm]
		$f_{\rm ice}$ & 0.4 & 0.087 \\
		& 0.8 &  0.128 \\
		& 1.2 &  0.149 \\
		& 1.6 &  0.170  \\
		& 2.0 &  0.176  \\   
		\hline
	\end{tabular}
	\label{tab:heightfeatures}
	\tablefoot{* indicates that we cannot measure the strength of the ice feature accurately due to the peak of the SED falling into the range of the water ice emission features.}
\end{table}

\subsection{Disk properties that do not change the ice features}

The disk inner and outer radius, within the limits we probe here, do not affect the water ice features. However, they are very instructive in narrowing down the region where the ice features predominantly form. The peak height of the $45~\mu$m feature is very constant between $0.11$ and $0.14$ throughout this series, that is they stay within $\pm\!17$\%, which is close to the error margin on the continuum placement (10\%). 

\begin{figure}[htb]
	\includegraphics[width=9.5cm]{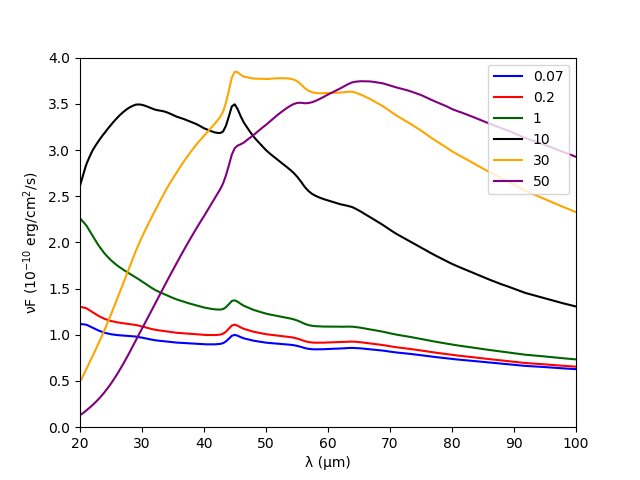} 
\caption{SED for the disk models with disk inner radius varying between 0.07 and 50~au.}
\label{fig:ZoomedSED_rin}
\end{figure}

Fig.~\ref{fig:ZoomedSED_rin} shows that while the continuum changes strongly with increasing disk inner radius from 0.07 to 50~au, the water ice features remain present. This indicates that the dominant contribution to the water ice features comes from beyond 50~au. This is also illustrated by Fig.~\ref{fig:2Dtemp_rin}: The ice reservoir (outlined by dashed white thick contour) stays unchanged for radii beyond $\sim\!50$~au. 

\begin{figure}[htb]
\includegraphics[width=9.5cm]{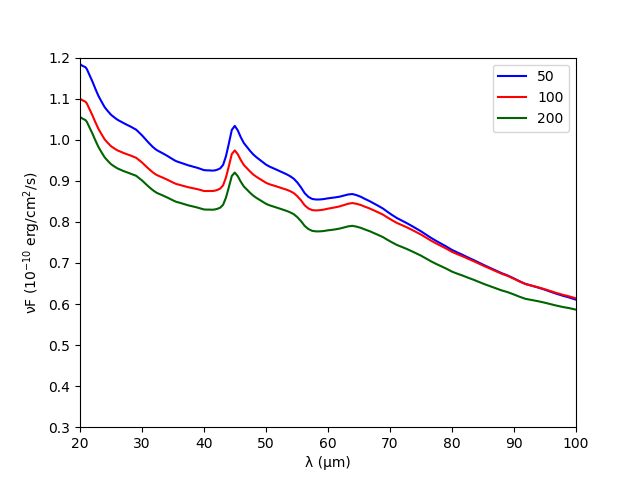}
\caption{SED for the disk models with disk tapering off radius varying between 50 and 200~au.}
\label{fig:ZoomedSED_rtap}
\end{figure}

Fig.~\ref{fig:ZoomedSED_rtap} shows that while the continuum changes strongly with increasing disk tapering off radius (50-200~au), the water ice features remain again very stable. Increasing the tapering off radius spreads the disk material to larger radii; however, this material further out is not substantially increasing the strength of the ice feature. This indicates that the dominant contribution to the water ice features comes from within $100$~au. The dust temperature within the ice reservoir beyond 100~au drops well below 50~K and hence the icy dust grains are not at the peak of their emissivity. Fig.~\ref{fig:2Dtemp_rtap} also shows that the ice reservoir touches the outer disk radius for large tapering off radii. However, the ice column density from those outer regions is likely too small to significantly contribute to the ice feature.

We use the turbulent $\alpha$ to parametrize dust settling \citep{Dubrulle1995}. The lower the turbulence, the stronger the dust settling in the outer disk. In the settled case, mostly the large grains settle towards the midplane, while the smaller micron-sized grains are less affected. These are the carriers of the far-IR ice features as demonstrated in the minimum grain size parameter series (see Sect.~\ref{sect:dustice-changefeature}). The water ice features change weakly with the level of turbulence. The ice feature becomes $\sim\!15$\% stronger for the lowest turbulence value compared with the highest one. This is largely due to the change in local continuum opacity pushing the $\tau\!\sim\!1$ lines at 45 and 63~$\mu$m closer to the disk midplane (see Fig.~\ref{fig:2Dtemp_alphaturb}). 

Of course this result is limited by the intrinsic assumption of settling being an equilibrium process and ice evaporation/re-condensation occurring on shorter timescales compared to possible vertical mixing processes.

\begin{figure}[htb]
	\includegraphics[width=9.5cm]{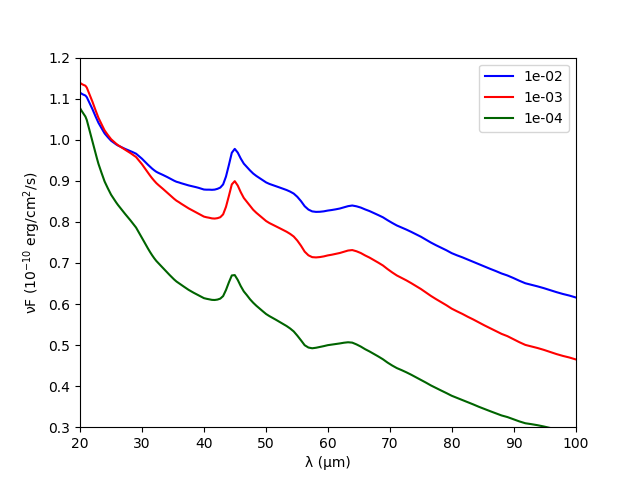}
\caption{SED for the disk models with different levels of turbulence $\alpha$ (see legend).}
\label{fig:ZoomedSED_alphaturb}
\end{figure}

\subsection{Dust/ice properties that change the ice features}
\label{sect:dustice-changefeature}

The 45 and $63~\mu$m water ice features are affected by the minimum grain size $a_{\rm min}$, the disk dust mass, the mass fraction of ice-to-refractory (so the thickness of the ice mantle) and also the thermal history of the water ice. In the following, we summarize the key results from those four model series.

\begin{figure}[htb]
	\includegraphics[width=9.5cm]{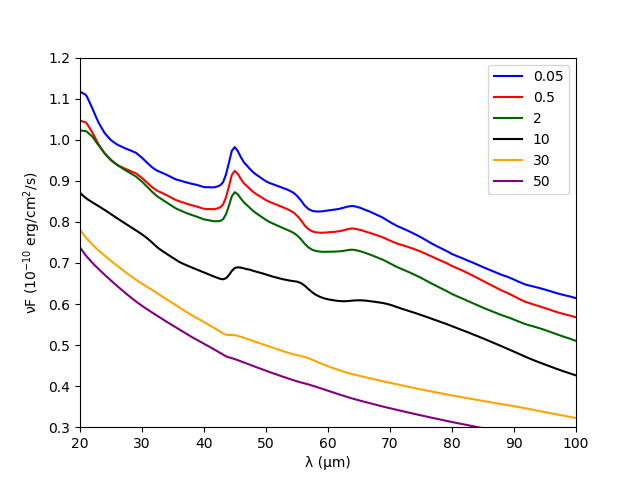}
\caption{SED for the disk models with minimum grain size varying between 0.05 and $50~\mu$m (see legend).}
\label{fig:ZoomedSED_amin}
\end{figure}

Fig.~\ref{fig:ZoomedSED_amin} shows that while the absolute level of the continuum changes strongly with increasing minimum grain size, the water ice features remain very stable up to $a_{\rm min}\!\sim\!10~\mu$m. Hence, we conclude that the dominant carrier of the far-IR water ice features are grains up to sizes of a few $\mu$m. This also explains the result in the previous section that dust settling is not affecting the strength of the 45 and $63~\mu$m water ice features. For a grain size distribution, where small grains are lacking, so $a_{\rm min}\!=\!10~\mu$m, the $45~\mu$m feature becomes a factor of three smaller than the reference model ($a_{\rm min}\!=\!0.05~\mu$m). 

The strength of the ice features correlates with dust mass until the disk reaches a mass of $\sim\!10^{-4}$~M$_\odot$. For higher dust masses, the ice features saturate and the feature strength levels off around 0.13 (Table~\ref{tab:heightfeatures}). Typical dust masses of class\,{\sc ii} disks around stars with less than 1~M$_\odot$ are below $2\,10^{-4}$~M$_\odot$ \citep[e.g.][]{Pascucci2016} and hence it should be possible to estimate the total ice mass in many young star forming regions.

\begin{figure}[htb]
	\includegraphics[width=9.5cm]{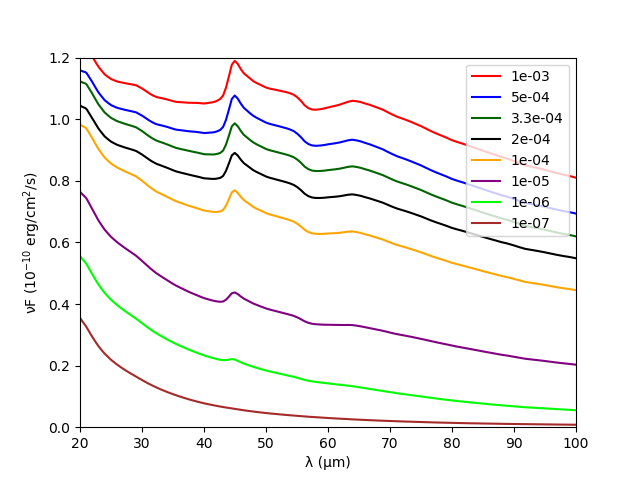}
\caption{SED for the disk models with disk dust masses varying between $10^{-3}$ and $10^{-7}$~M$_\odot$ (see legend).}
\label{fig:ZoomedSED_Mdust}
\end{figure}

The ice fraction produces also a strong effect: The thicker the ice mantle, the larger the peak strength. However, at high ice fractions of $>\!1.6$, the peak strength saturates. Adding more ice to the mantle does not contribute anymore to the emission; the ice becomes `optically thick'. Overall, an increase in the thickness of the ice mantle by a factor of $1.3$ causes an increase in peak strength by a factor of approximately two ($f_{\rm ice}\!=\!0.4$ with respect to $1.6$).

We would like to note here that we assume in the above model series the ice to be in crystalline shape (i.e.\ formed at temperatures above 100~K), causing prominent features. This assumption is plausible given the results from \citet{Min2016b}. However, if the ice in the disk is formed and still located at temperatures below 100~K, a significant fraction could be in amorphous form, which is more difficult to observe.

\begin{figure}[htb]
	\includegraphics[width=9.5cm]{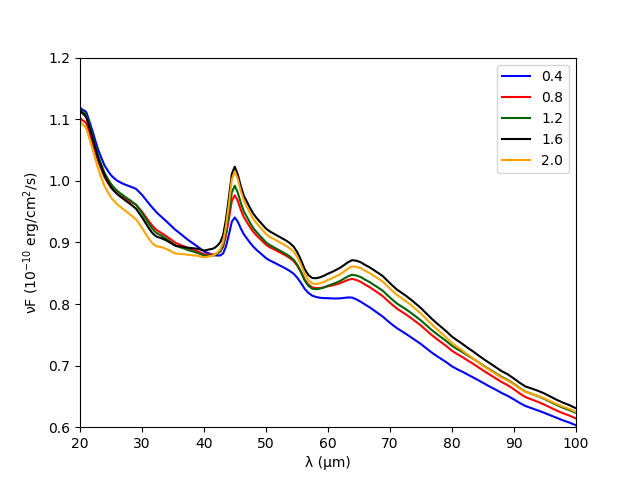}
\caption{SED for the disk models with different water ice fraction $f_{\rm ice}$ on the grains ($f_{\rm ice}\!=\!2$ means that the models contains two times the dust mass added in ice mantles). }
\label{fig:ZoomedSED_ice-fraction}
\end{figure}

\begin{figure}[htb]
	\includegraphics[width=9.5cm]{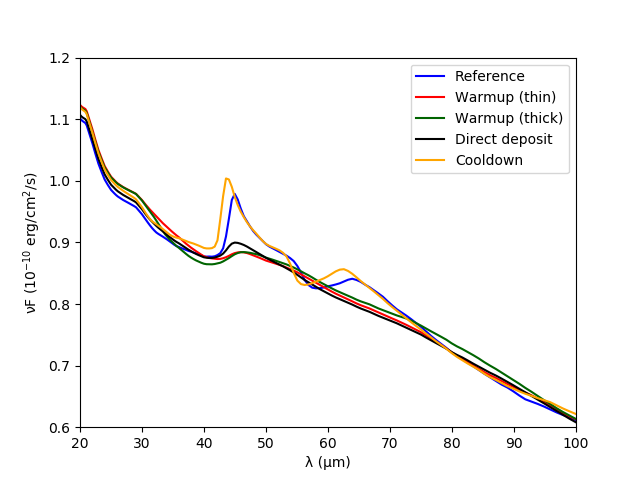}
\caption{SED for the disk models with different water ice `thermal histories'.}
\label{fig:ZoomedSED_ice-scenarios}
\end{figure}

Figure~\ref{fig:ZoomedSED_ice-scenarios} shows that the different thermal histories of water ice will lead to clearly different feature strength and shape. Table \ref{tab:locationfeatures} shows the peak wavelength of both water ice features for these series. The $63~\mu$m feature shows no clearly defined peak for the warmup and direct deposit scenarios. 

Most interestingly, in the cooldown series, we see a significant shift of the peak of the $45~\mu$m ice feature to shorter wavelengths ($43.55~\mu$m) with respect to the reference series with opacities at 140~K ($44.96~\mu$m). This indicates that a large fraction of the ice feature originates rather from $\sim\!70$~K ice \citep[see Table~1 of][]{Smith1994}. The $63~\mu$m feature also shows a shift corresponding to $\sim\!90$~K ($62.6~\mu$m). In the reference model, temperatures of 70-90~K correspond to temperatures at the snow line surface at a distance of 10-20~au. 

In the direct deposit and warm-up series, the water ice features are more indicative of amorphous ice. Also, their peaks shift to longer wavelengths ($45.93~\mu$m and $46.42~\mu$m) with respect to the reference model, again indicating on average low ice temperatures in the emitting region ($< \! 110$~K) in agreement with their amorphous nature and an origin beyond 6~au. 

\begin{table}[htb]
	\caption{Peak wavelengths position of the water ice features for the different thermal history scenarios.}
	\begin{tabular}{lll}
		\hline
		\hline\\[-3mm]
		& 45~$\mu$m  & 63~$\mu$m  \\
		\hline\\[-3mm]
		Reference      &  44.96  &  63.91 \\
		Warmup (thin)  &  45.93  &  - \\
		Warmup (thick) &  46.42  &  - \\
		Direct deposit &  44.96  &  - \\
		Cooldown       &  43.55   &  62.56 \\
		\hline		
	\end{tabular}
	\label{tab:locationfeatures}
\end{table}

This model series is the only one where we include proper temperature dependent ice opacities. It shows that our approach to find the spatial origin of these water ice features using the model series with varying inner and tapering-off radii might be too simple. For example, the emitting region of the ice features could change in the model series with increasing $R_{\rm in}$, and an increasing surface area could compensate for the emission becoming weaker, thus leading to the apparent `constancy' of the far-IR water ice features over a wide range of disk geometries. The disk model series with water ice histories may provide here a more reliable estimate of the main emitting region through the measurement of the $45~\mu$m ice feature peak.

\section{Future observations}
\label{sec:future}

\subsection{SOFIA/HIRMES}

HIRMES was selected as a third generation instrument for the SOFIA
 Observatory. The SOFIA Observatory consists of a 2.5~m telescope mounted
 inside a modified Boeing~747SP operating at altitudes from 12-14~km.
 HIRMES is an infrared spectrograph operating between 25 and 122~$\mu$m
 with a spectral resolving power of 350-650 in low resolution mode and
 $5\,10^4-10^5$ in high resolution mode. HIRMES will be commissioned on
 SOFIA in early to mid 2019. For the low resolution mode, rebinning to a resolution of 100, 
 the expected sensitivity is $\sim\!0.04$~Jy (1$\sigma$ in 1~hr; Klaus Pontoppidan, private communication)).

\begin{figure}[htb]
	\includegraphics[width=9.cm]{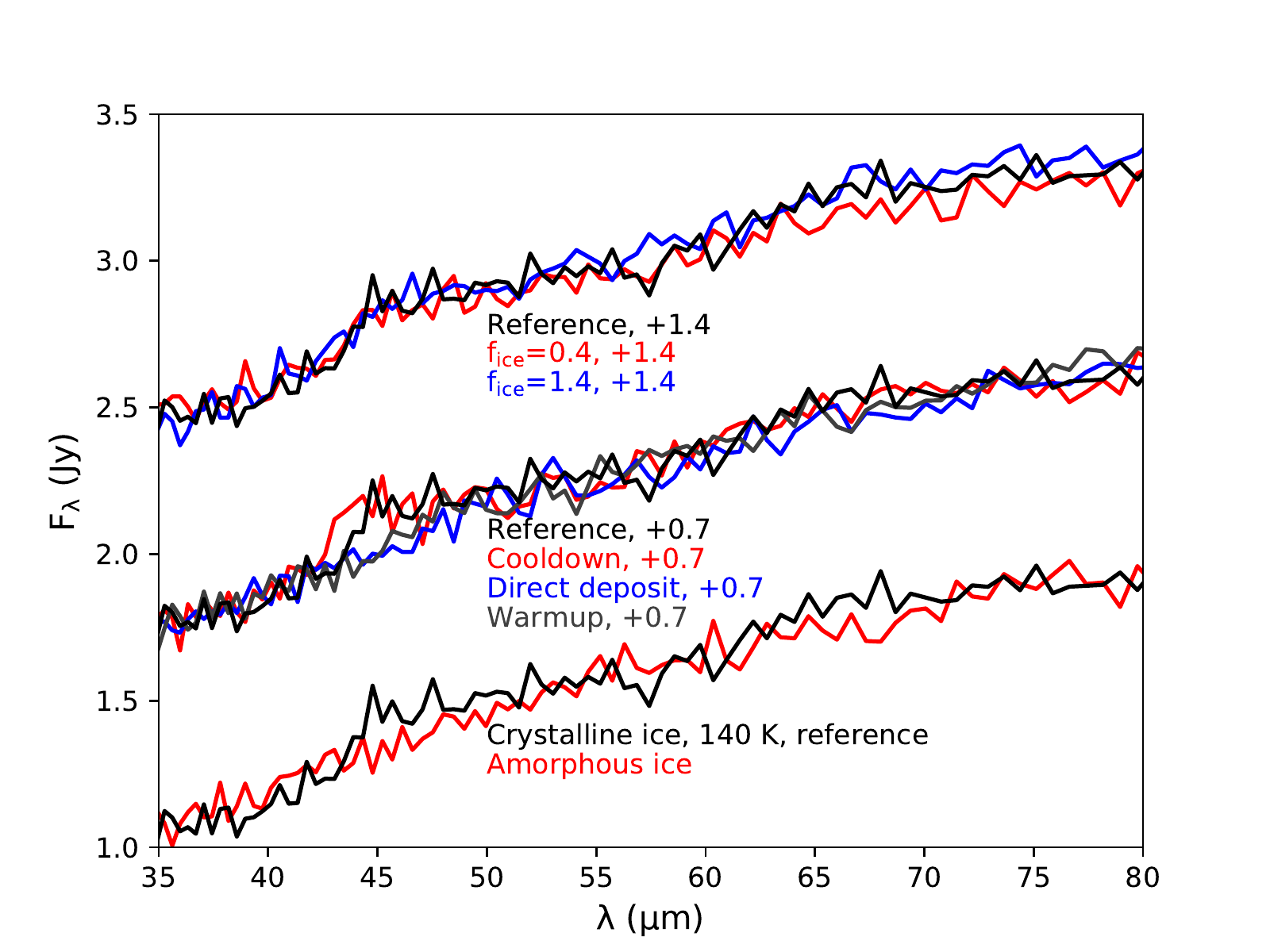}
\caption{Simulated HIRMES spectra for several ice model series: crystalline ice and amorphous ice, different water ice `thermal histories' shifted upwards by 0.7~Jy ($+0.7$) and different ice thicknesses shifted upwards by 1.4~Jy ($+1.4$). The exposure time is 1~hr and the resolution is $R\!=\!100$.}
\label{fig:HIRMESice-composite}
\end{figure}

Figure~\ref{fig:HIRMESice-composite} shows the noise level that HIRMES will achieve in a 1 hr exposure for various of our model series using $R\!=\!100$. Even though it could provide detections in the case of brighter sources and/or for particularly strong features (e.g. very high ice fraction, crystalline ice at 140~K), for most of the model simulations of this paper detailed studies of the ice properties and thermal history will be challenging with a resolution of $R\!=\!100$. Given the current sensitivities and our model predictions, ice studies of T Tauri disks with HIRMES will require long exposure times, for typical T Tauri continuum fluxes (1~Jy and lower) even beyond 1 hour. 

\subsection{SPICA/SAFARI}

The SPace Infrared telescope for Cosmology and Astrophysics (SPICA) is a joint Japanese/European 2.5~m cooled telescope proposed for ESAs M5 call. The SAFARI instrument on board will allow spectroscopy between 34 and 230~$\mu$m at a resolution of 300-11000. In low resolution mode, it will reach 0.31, 0.45 and 0.72~mJy (5$\sigma$ in 1~hr) in the SW (34-56~$\mu$m), MW (54-89~$\mu$m) and LW (87-143~$\mu$m) band respectively.

\begin{figure}[htb]
	\includegraphics[width=9.cm]{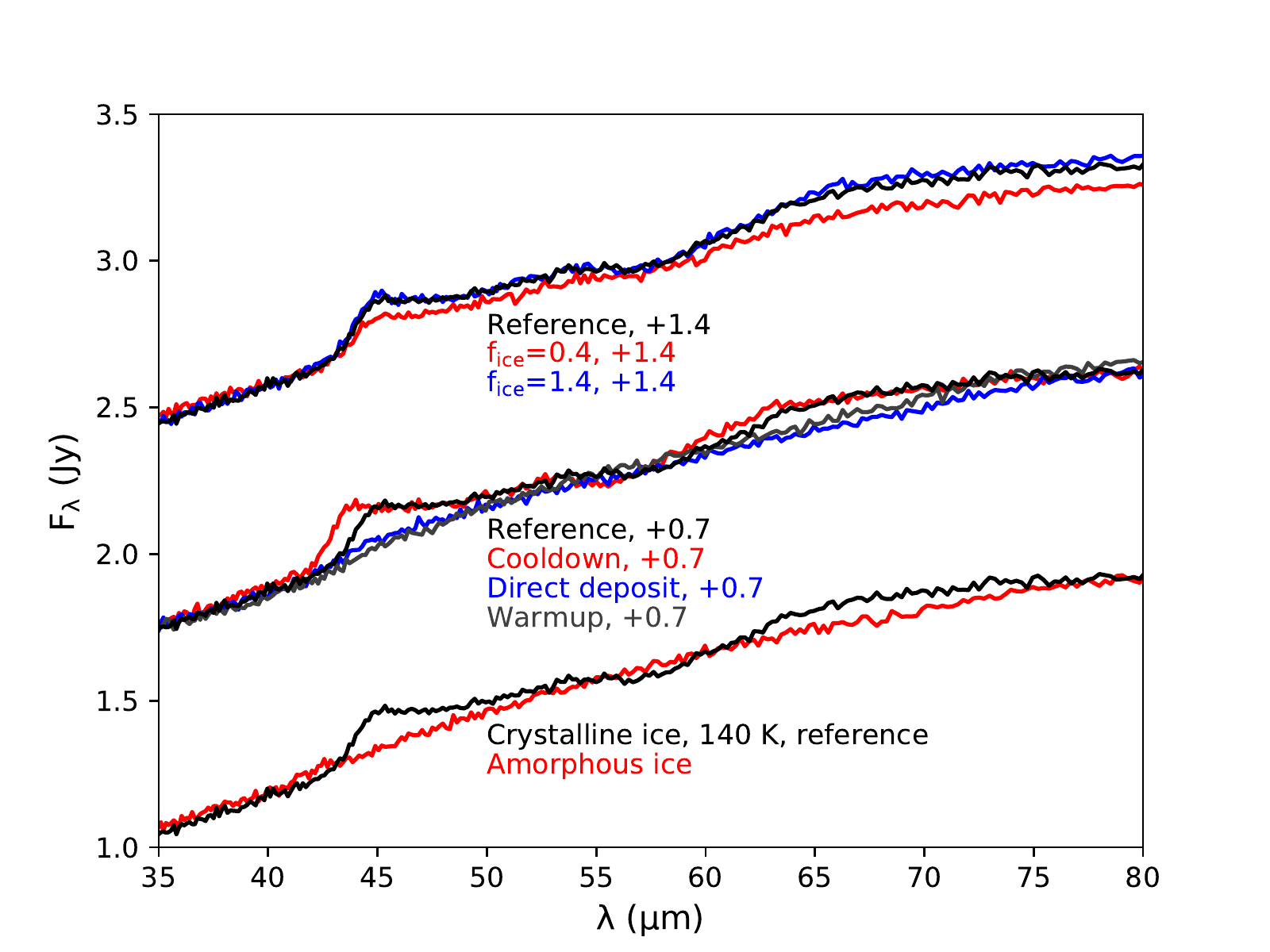}
\caption{Simulated SAFARI spectra for several ice model series: crystalline ice and amorphous ice, different water ice `thermal histories' shifted upwards by 0.7~Jy ($+0.7$) and different ice thicknesses shifted upwards by 1.4~Jy ($+1.4$). The exposure time is 10~min and the resolution is $R\!=\!300$.}
\label{fig:SAFARIice-composite}
\end{figure}

Figure~\ref{fig:SAFARIice-composite} shows the noise level that SAFARI will achieve in a short 10~min exposure for various of our model series using $R\!=\!300$. The \mbox{SAFARI} instrument with its high sensitivity will be able to distinguish between cooldown and direct deposit/warmup ice histories, but also ice thicknesses, i.e.\ ice fractions. The short exposure times (10~min) will enable ice surveys of hundreds of T~Tauri disks with the SPICA mission. The most important limitation will be the baseline stability and calibration of the instrument. For detailed ice characterization studies, we require a stability of better than a few \% across the 35-100~$\mu$m spectral range.

Long exposures (1~hr) can provide more reliable ice fraction measurements. However, the distinction between direct deposit and warmup scenarios, and the detection of amorphous ice will remain difficult and hinges mostly on a reliable continuum placement due to the broadness of the weak features. 

\section{Discussion} 
\label{sec:discussion}

The constant nature of the far-IR water ice features against physical disk parameters such as disk sizes and settling makes these features a robust probe of the grain and ice history in disks.

A closely related water ice feature study is \citet{McClure2015}. One of their Herschel observations, DO~Tau, shows far-IR continuum fluxes very close to our reference model. The peak strength of the detected feature at $63~\mu$m is $\sim\!0.2$, at least a factor of two stronger than the feature we typically see using crystalline water ice in our reference model. The other water ice detections they report are even stronger. On the other hand, the weak features predicted from our `thermal history' modeling series, would have been undetectable given the typical PACS noise level. The only way in our modeling approach to enhance the strength of the $63~\mu$m feature to a level that would have been detectable by PACS, would be to neglect the photodesorption and have the water ice reservoir extend high up in the disk. It is however important to keep in mind that the continuum placement with PACS was very difficult due to the pointing stability and the data was obtained in range spectroscopy mode (51-73~$\mu$m) thus only covering the central part of the feature without much neighboring continuum.

The `thermal history' modeling series illustrates the power of having access to the peak position of the $45~\mu$m ice feature, the strength of the peak as well as the shape of the $45$ and $63~\mu$m feature complex. If the ice is purely amorphous, the $63~\mu$m feature will be hard to distinguish from the continuum due to its very broad nature. However, the shorter wavelength feature remains in that case an important diagnostic and the combination of peak wavelength and strength has the potential to discriminate between in-situ formation and a scenario where the ice has been formed cold either in the precursor molecular cloud or in the outer disk and later distributed more widely through radial migration processes. The opposite process, a formation close to the midplane snow line and later migration outwards leaves the ice in a crystalline state, which is clearly detectable in both far-IR water ice features. 

The next step should be to couple detailed grain growth/migration models with the self-consistent snow line and dust radiative transfer approach. The optical depth of the disk will likely change if the grain size distribution varies with position in the disk. This could affect the appearance and strength of the water ice features. A step in this direction is the dust settling series in which the height of the far-IR continuum of the disk changes substantially. Similar changes can be expected for a varying radial distribution driven by dust migration. Once the smaller dust grains change their spatial distribution, also the UV/optical $\tau\!\sim\!1$ line will change and this can also change the location of the snow line in the disk surface. Dust drift models suggest a high efficiency in depleting small $\mu$m-sized grains from a wide region around $\sim\!80$~au \citep[region\,{\sc iv} in Fig.~2 of][]{Birnstiel2015}. This is the region we identify here as the main contributor to the water ice thermal emission feature. Leaving this region devoid of particle sizes that can produce the thermal ice emission features (grains $\lesssim\!10~\mu$m) would have an effect on the strength of the feature.

It is interesting to note that the absence of a water ice feature in the spectrum does not imply the absence of an extended water ice reservoir. This can be caused by the absence of grains smaller than $10~\mu$m in the outer parts of the disk ($\gtrsim\!30$~au). The disk model with $a_{\min}\!=\!30~\mu$m shows no features in the SED, but still shows a significant ice reservoir. 


\section{Conclusions} 
\label{sec:conclusion}

This paper studies for the first time the far-IR water ice features in a self-consistent modeling approach to set the location of the water snowline together with consistent dust/ice opacities and dust disk thermal structure. We demonstrate that photodesorption of water ice is key in the quantitative interpretation of the peak strengths of the 45 and $63~\mu$m features. The water ice features originate from grains smaller than $\sim\!10~\mu$m.

The water ice features originate predominantly at the surface snow line of the disk between 10 and 100~au. As such they are very robust against several disk parameters such as inner and outer radii, within limits the grain size distribution and dust settling. However, the feature strength scales with disk dust mass in the range of typical T Tauri disks. The features also
change substantially with the thickness of the ice mantle, although the strength levels off for mantles much thicker than 2.5 times the original grain sizes.

Most interestingly, the features do carry an imprint of the thermal history of the ice and thus can distinguish between a warmup scenario for the water ice (water ice is formed in cold, optically thick regions and then moves to warmer areas in the disk) and a cooldown scenario (water ice is formed in warm areas close to the midplane snowline and then moves outwards to colder areas in the disk). The direct deposit scenario (ice stays where it was formed locally in the disk) resembles qualitatively the warmup scenario and while the peak strength change of the 45~$\mu$m feature may not allow a discrimination, the shift of the peak wavelengths to shorter values might be a discriminator.

Given the results of our ice model series, simulations using the sensitivities for the planned SOFIA/HIRMES instrument show that the low resolution as well as the sensitivity will severely limit detailed ice studies --- beyond simple detection --- in T Tauri disks at distance of typical star forming regions. The $45~\mu$m ice feature is strong enough to be detected if it originates from warm grains with thick ice mantles that are crystalline in nature. Similar simulations for the SPICA/SAFARI instrument show that both the differences in peak strengths as well as the wavelength shifts in the peak position can be easily measured for samples of several hundred T Tauri disks (exposure times of 10~min). At low spectral resolution ($R\!=\!300$) weak flat ice features at $45~\mu$m (e.g.\ amorphous, direct deposit, warmup of ice) require a good continuum characterization in deep surveys.

\begin{acknowledgements}
IK and MM acknowledge funding from the EU FP7-2011 under Grant Agreement no. 284405. LK  is  supported  by  a  grant from the Netherlands Research School for Astronomy (NOVA). We thank the anonymous referee for comments that improved the clarity of the paper and the suggestion to include the dust mass series.
\end{acknowledgements}

\newpage
\clearpage

\begin{appendix}

\section{SEDs and disk structures}
\label{App:modelseries}

\subsection{Disk inner radius and tapering off radius}
\label{App:sect:rin:rtap}

\begin{figure*}[tbh]
	\includegraphics[width=6.cm]{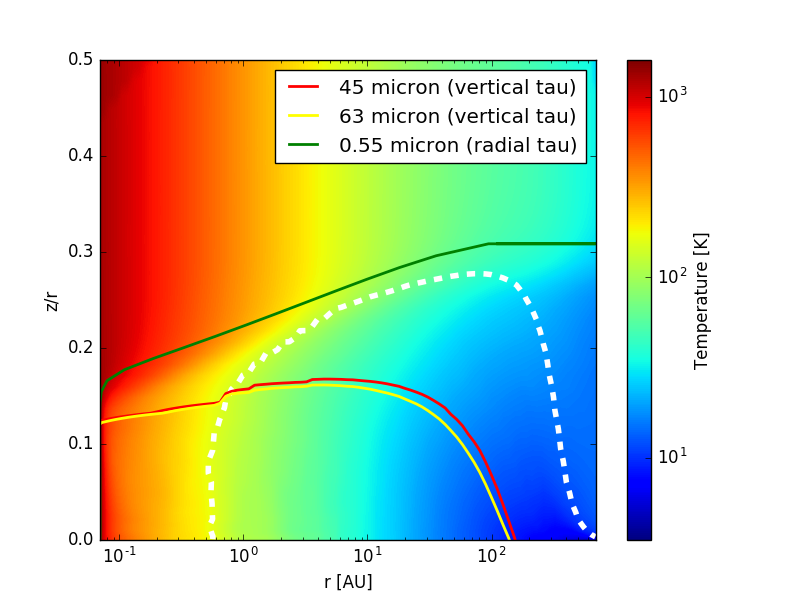} 
	\includegraphics[width=6.cm]{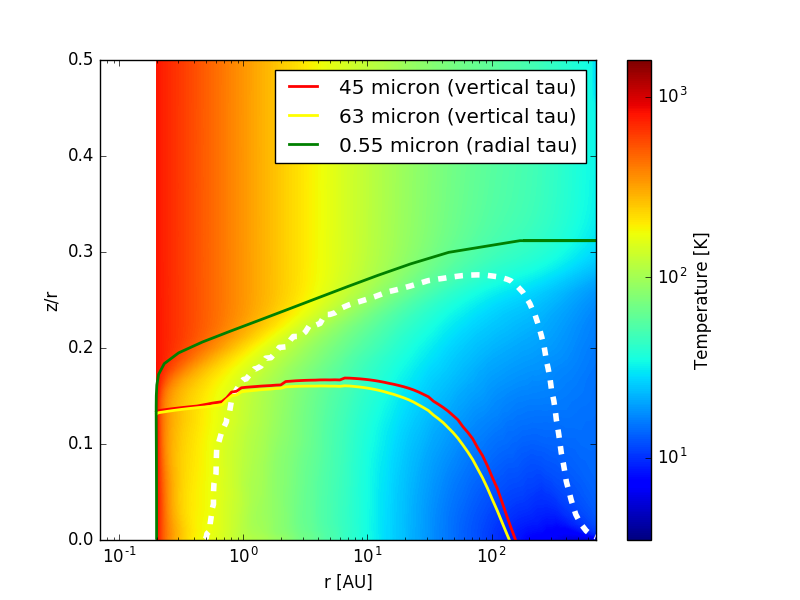}
	\includegraphics[width=6.cm]{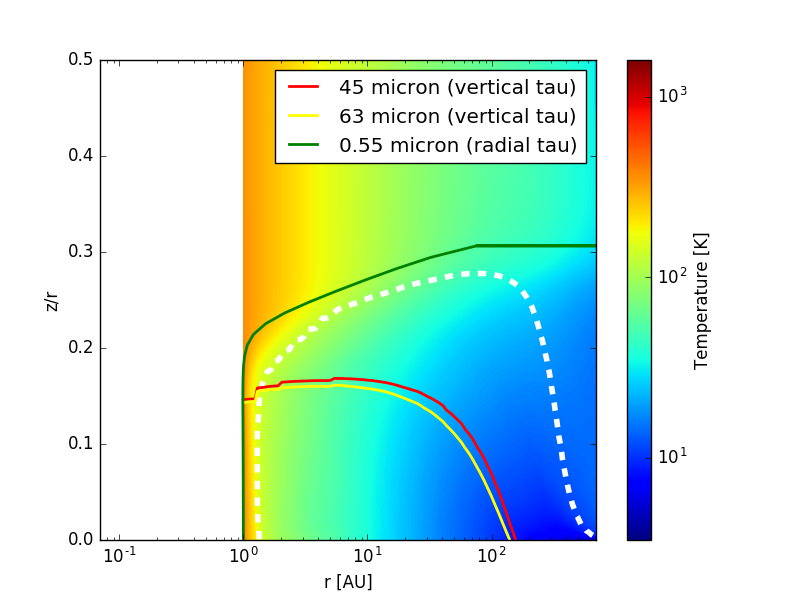} 
	\includegraphics[width=6.cm]{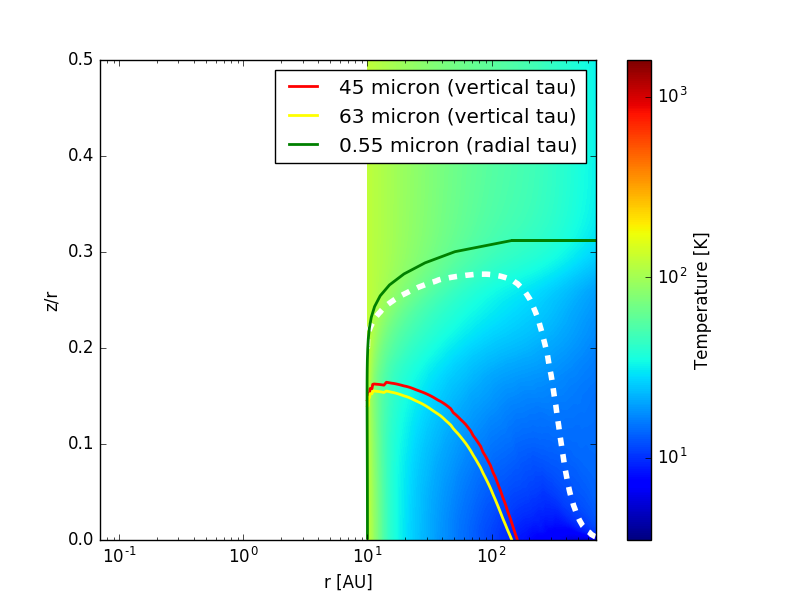}
	\includegraphics[width=6.cm]{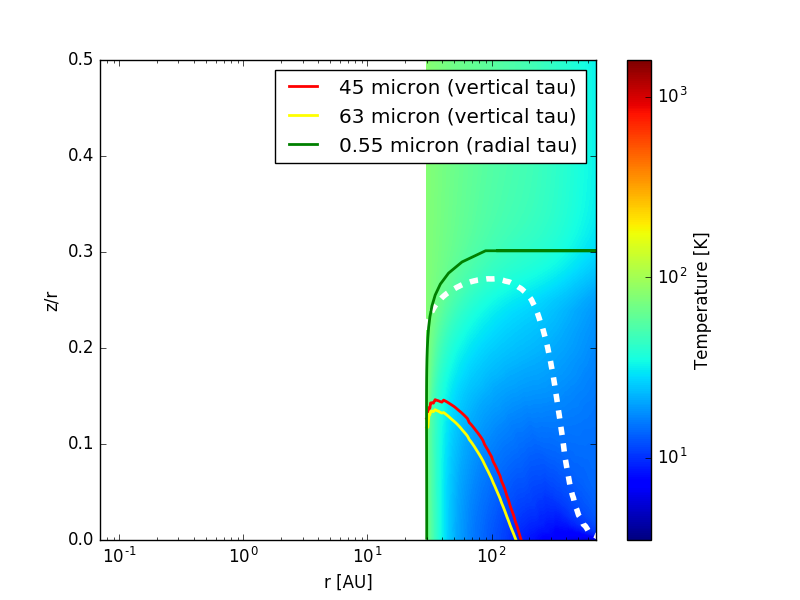} 
	\includegraphics[width=6.cm]{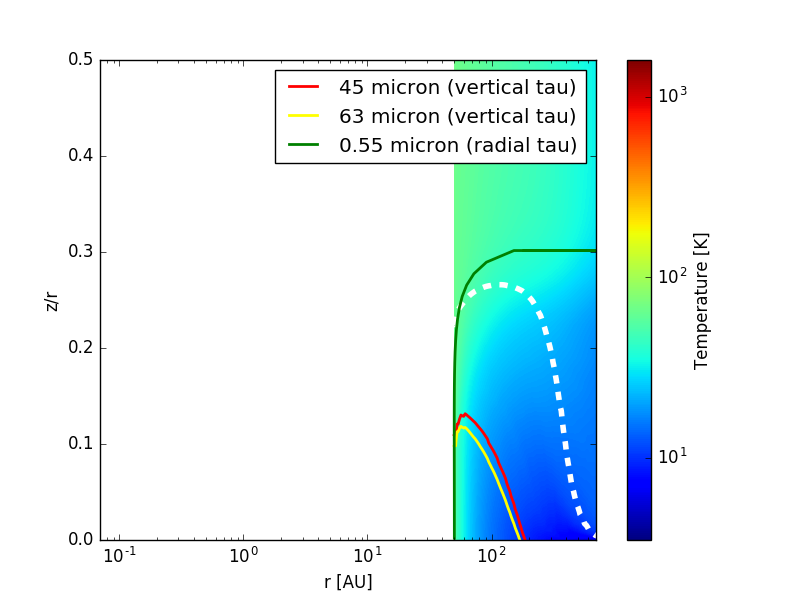} 
\caption{The 2D dust temperature distribution in the model series where the disk inner radius changes from 0.07 to 50~au. The contour lines have the same meaning as in Fig.~\ref{fig:2Dtemp_referencemodel}.}
\label{fig:2Dtemp_rin}
\end{figure*}

\begin{figure*}[!htb]
	\includegraphics[width=6.cm]{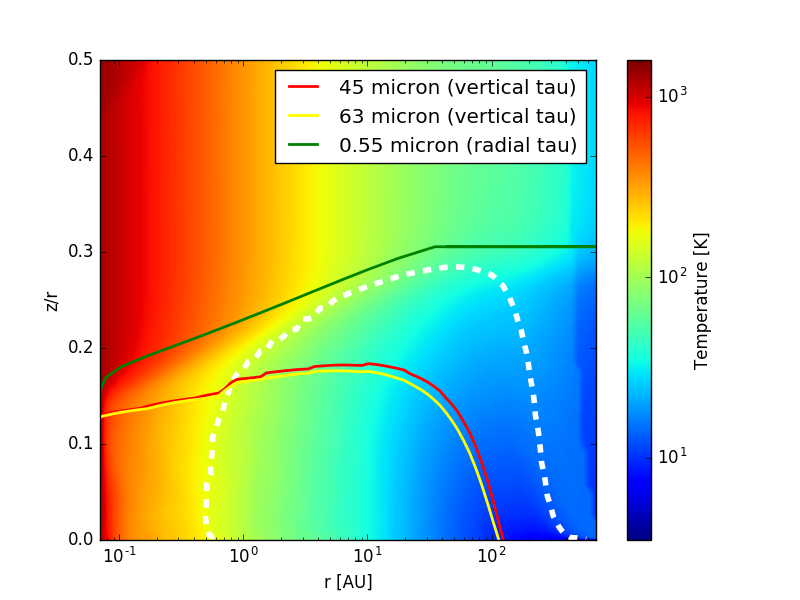} 
	\includegraphics[width=6.cm]{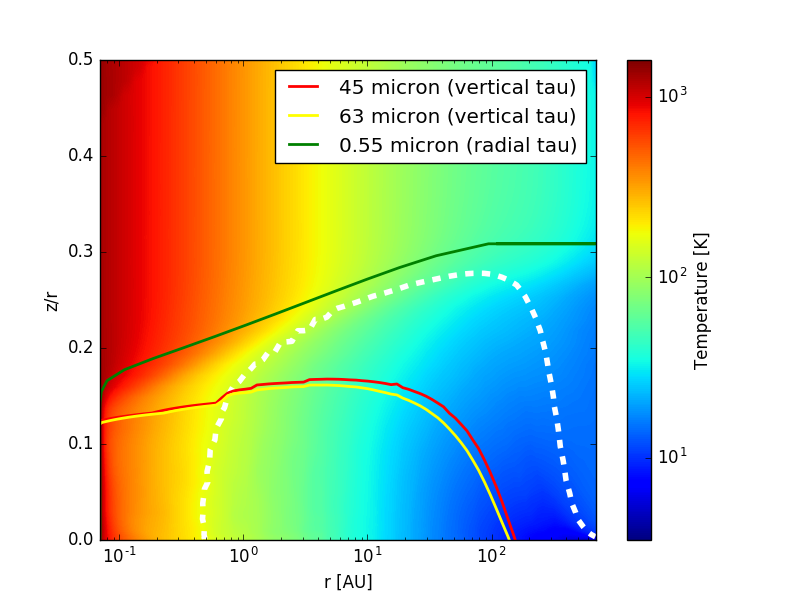} 
	\includegraphics[width=6.cm]{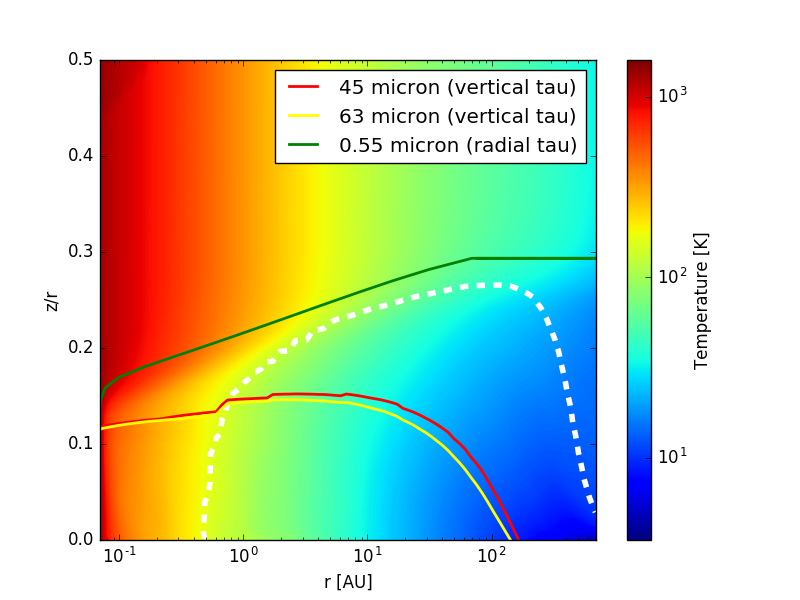}
\caption{The 2D dust temperature distribution in the model series where the disk tapering off radius changes from 50 to 200~au. The contour lines have the same meaning as in Fig.~\ref{fig:2Dtemp_referencemodel}.}
\label{fig:2Dtemp_rtap}
\end{figure*}

Fig.~\ref{fig:2Dtemp_rin} and Fig.~\ref{fig:2Dtemp_rtap} illustrate the temperature distribution in the disk models with varying inner radius and tapering off radius. To assess where the water ice features originate, we also show the snow line and the vertical optical depth $\tau\!=\!1$ contours for 45 and $63~\mu$m and the radial optical depth $\tau\!=\!1$ at $0.55~\mu$m.

\subsection{Dust settling}
\label{App:sect:alphaturb}

\begin{figure*}[htb]
	\includegraphics[width=6cm]{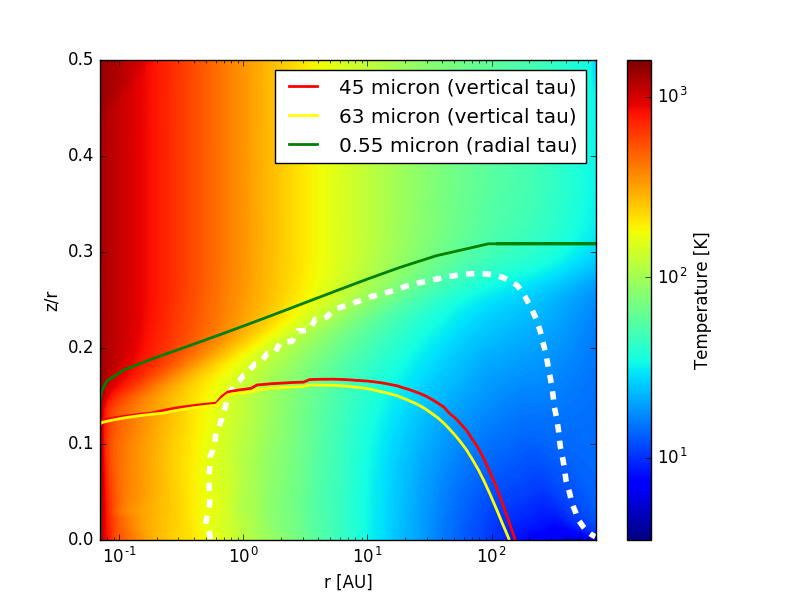} 
	\includegraphics[width=6cm]{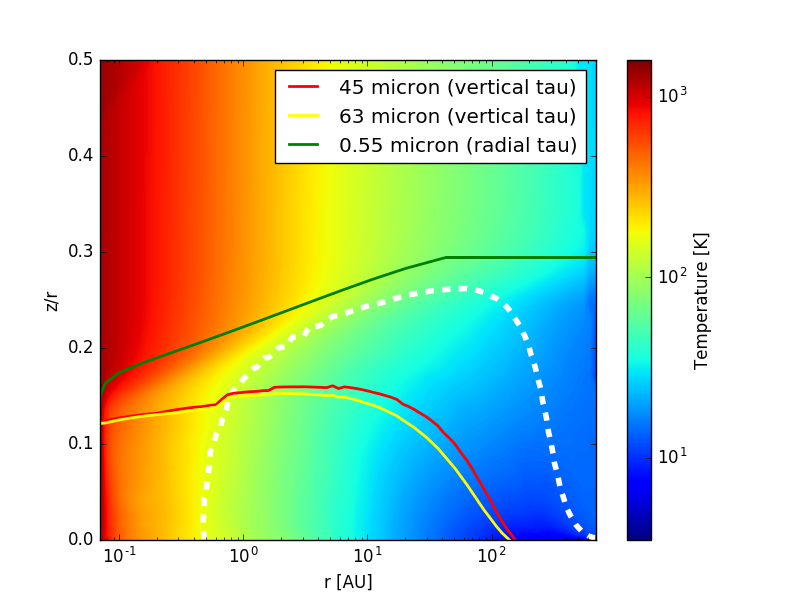} 
	\includegraphics[width=6cm]{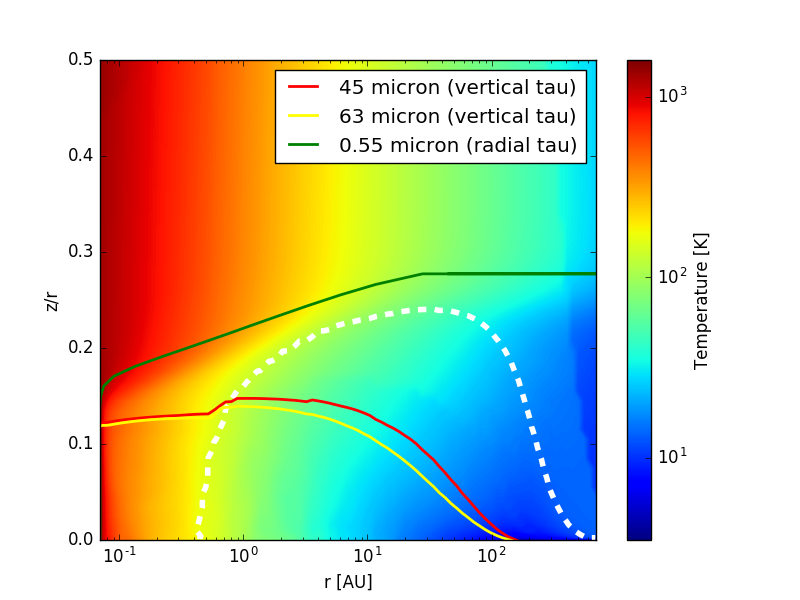} 
\caption{The 2D dust temperature distribution in the model series where the turbulence decreases from 0.01 to $10^{-4}$. The contour lines have the same meaning as in Fig.~\ref{fig:2Dtemp_referencemodel}.}
\label{fig:2Dtemp_alphaturb}
\end{figure*}

Fig.~\ref{fig:2Dtemp_alphaturb} shows the 2D temperature distribution for disk models with various degrees of settling ($\alpha_{\rm turb}\!=10^{-2}$ to $10^{-4}$). The change of opacities is strongest in the outer disk and hence temperature changes there are largest. Lower values for $\alpha_{\rm turb}$ lead to strong dust settling towards the midplane, thus reducing the opacities in the upper/outer layers. This shifts the snow line to lower heights (from $\sim\!0.3$ at 100 au for $\alpha_{\rm turb}\!=\!10^{-2}$ to $\sim\!0.25$ for $\alpha_{\rm turb}\!=\!10^{-4}$) and produces a larger water vapor reservoir in the outer disk.

\subsection{Minimum grain size}
\label{App:sect:amin}

\begin{figure*}[htb]
	\includegraphics[width=6cm]{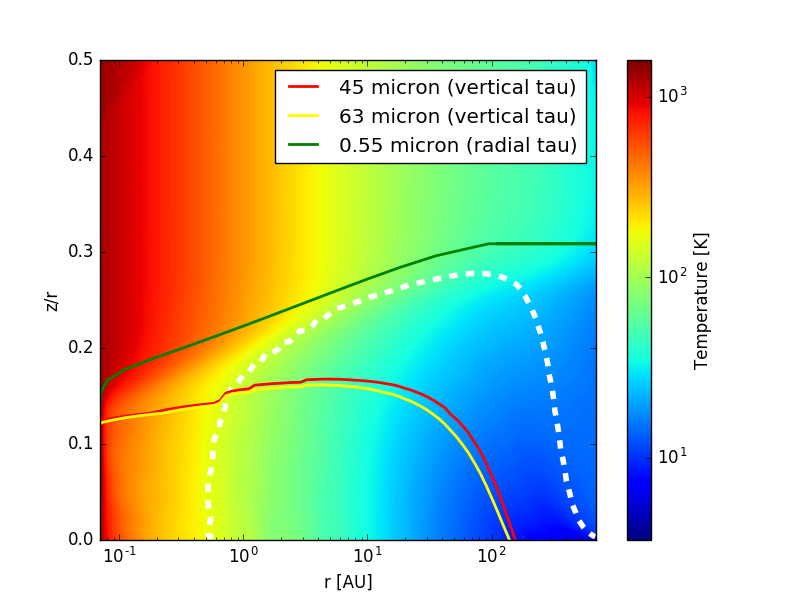} 
	\includegraphics[width=6cm]{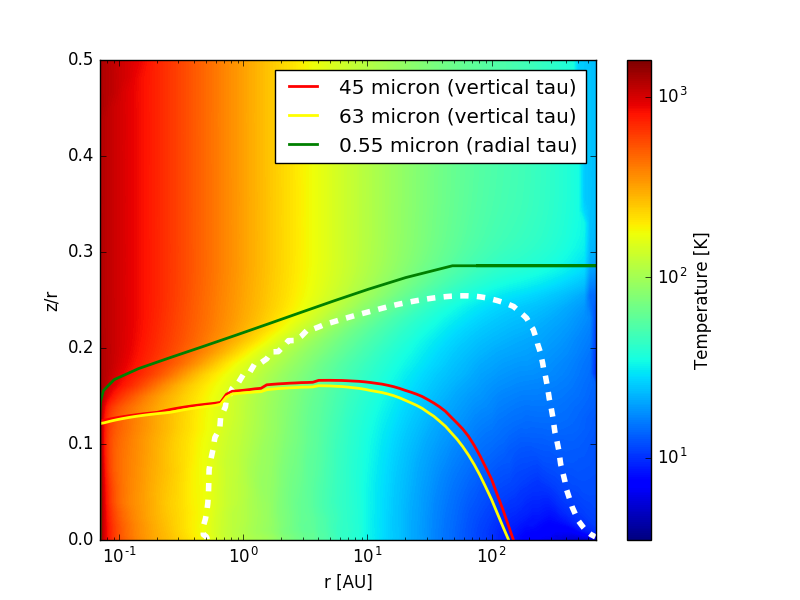} 
	\includegraphics[width=6cm]{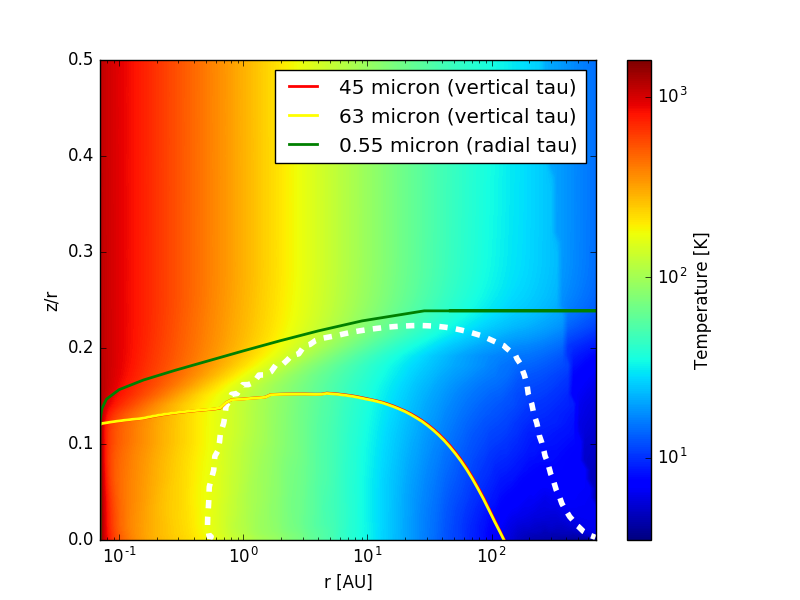} 
	\includegraphics[width=6cm]{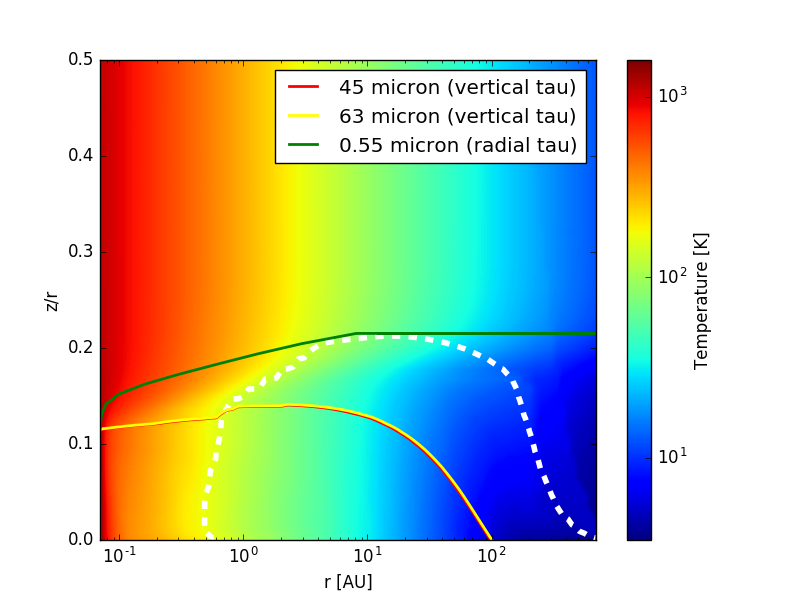} 
	\includegraphics[width=6cm]{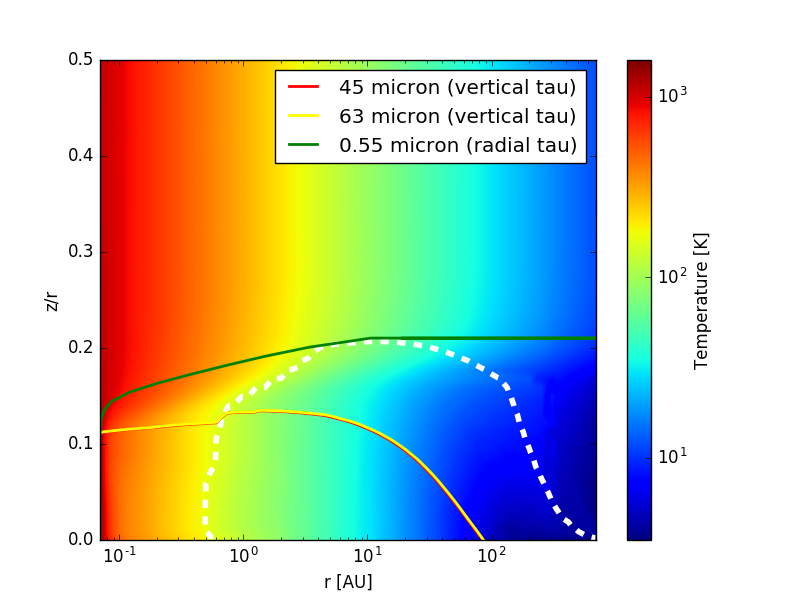} 
\caption{The 2D dust temperature distribution in the model series where the minimum grain size increases from 0.05 to $50~\mu$m. The contour lines have the same meaning as in Fig.~\ref{fig:2Dtemp_referencemodel}.}
\label{fig:2Dtemp_amin}
\end{figure*}

When we change the minimum grain size from 0.05 to $50~\mu$m, the optical depth in the disk changes dramatically as illustrated in  Fig.~\ref{fig:2Dtemp_amin}. Removing the smaller grains lowers the opacity and the largest impact is seen in the surface layers, especially in the change of height at which the disk becomes optically thick (green contour, $\tau\!=\!1$ at $0.55~\mu$m). Subsequently, also the snow line moves closer to the midplane ($\sim\!0.3$ at 100~au for $a_{\rm min}\!=\!0.05~\mu$m grains to $\sim\!0.2$ for $a_{\rm min}\!=\!50~\mu$m grains). 

If the disk surfaces are strongly depleted in small grains, there is an increasing chance of observing the ice feature also in scattered light; the green radial optical depth line touches the snow line, i.e.\ optical stellar light can be scattered off icy grains even at large distances from the star (out to several 10~au).

\subsection{Disk dust mass}
\label{App:sect:Mdust}

\begin{figure*}[htb]
	\includegraphics[width=6cm]{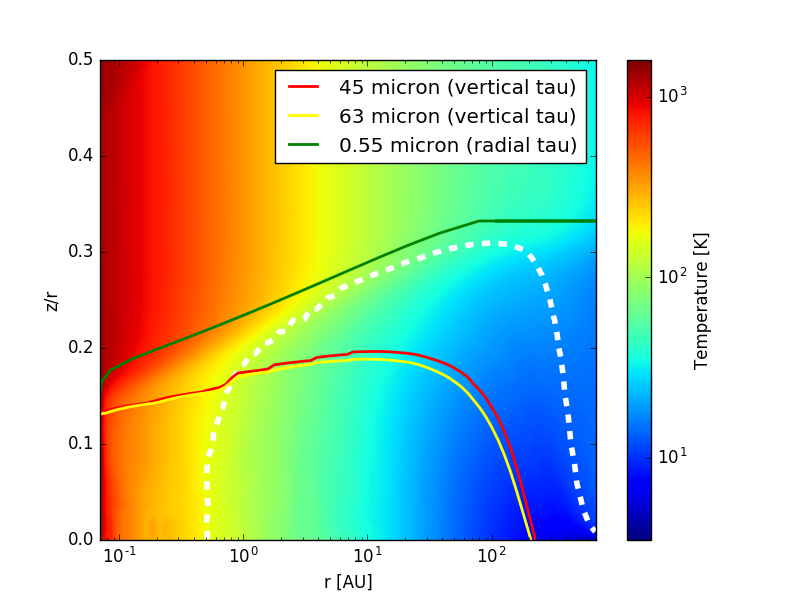} 
	\includegraphics[width=6cm]{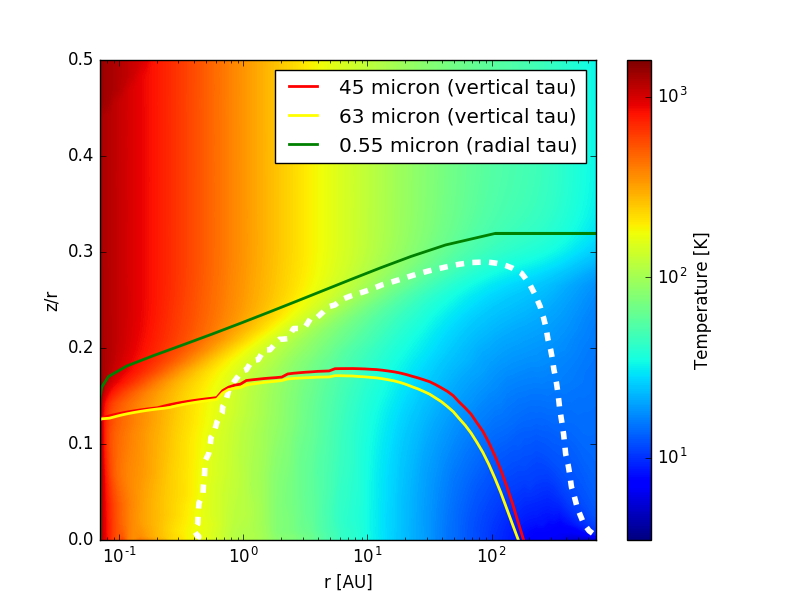} 
	\includegraphics[width=6cm]{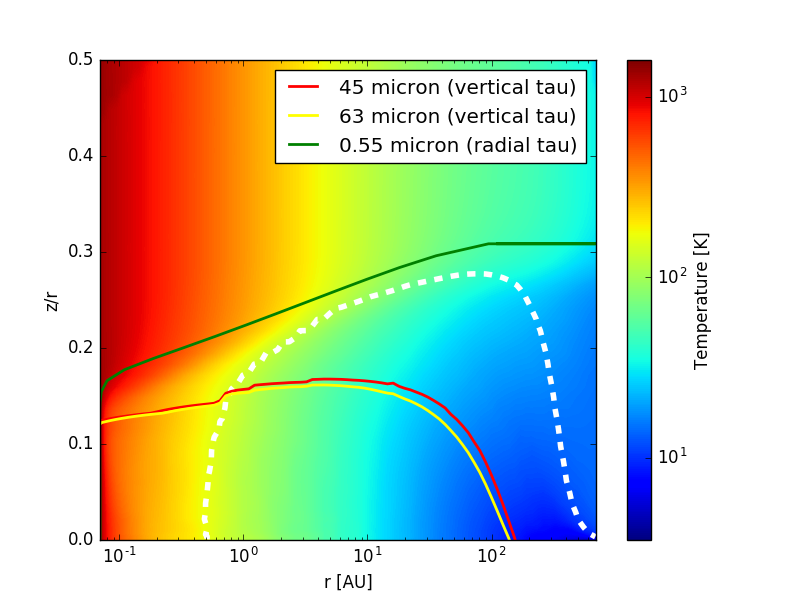} 
	\includegraphics[width=6cm]{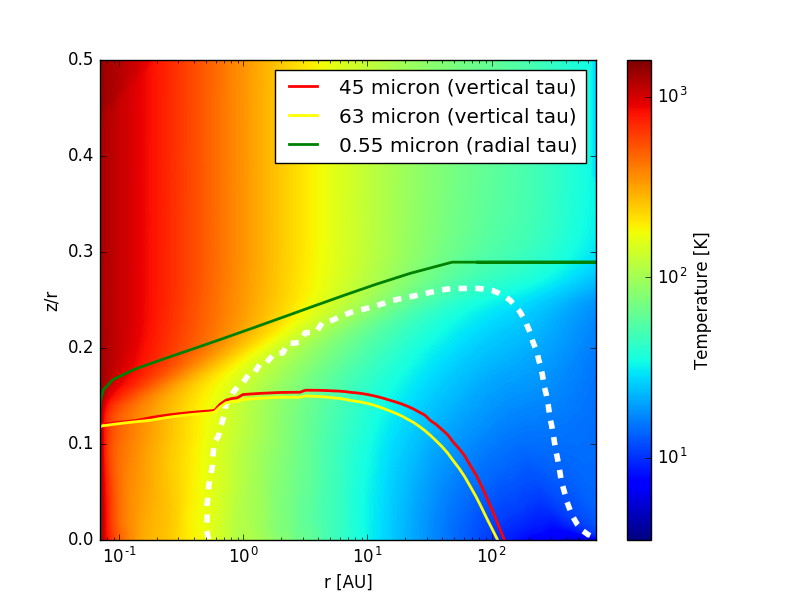} 
	\includegraphics[width=6cm]{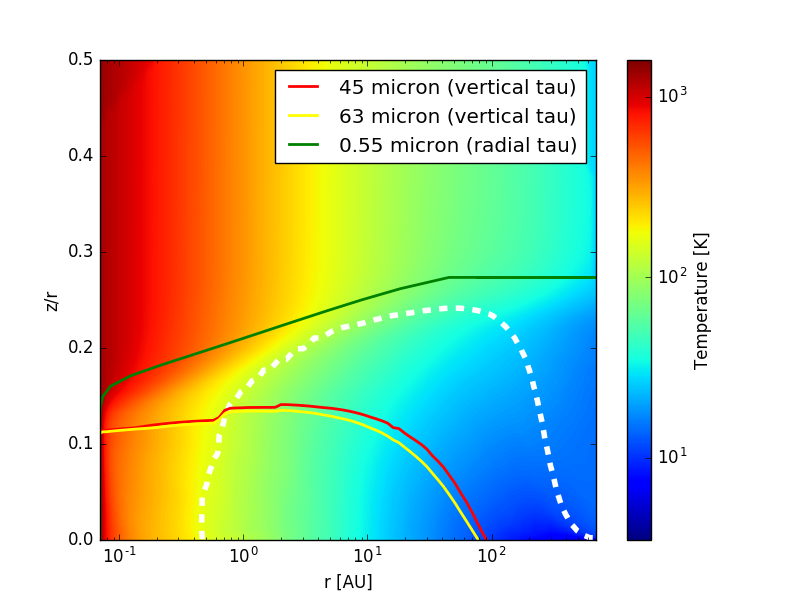} 
	\includegraphics[width=6cm]{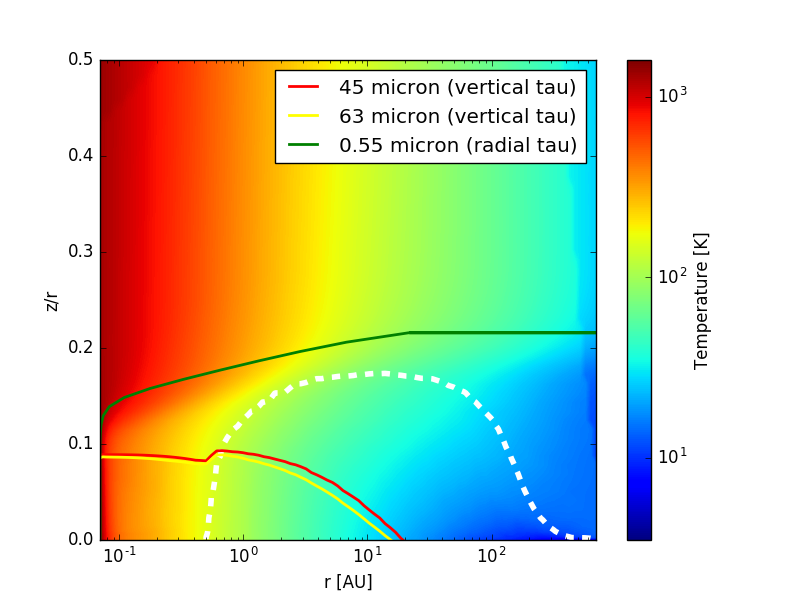} 
	\includegraphics[width=6cm]{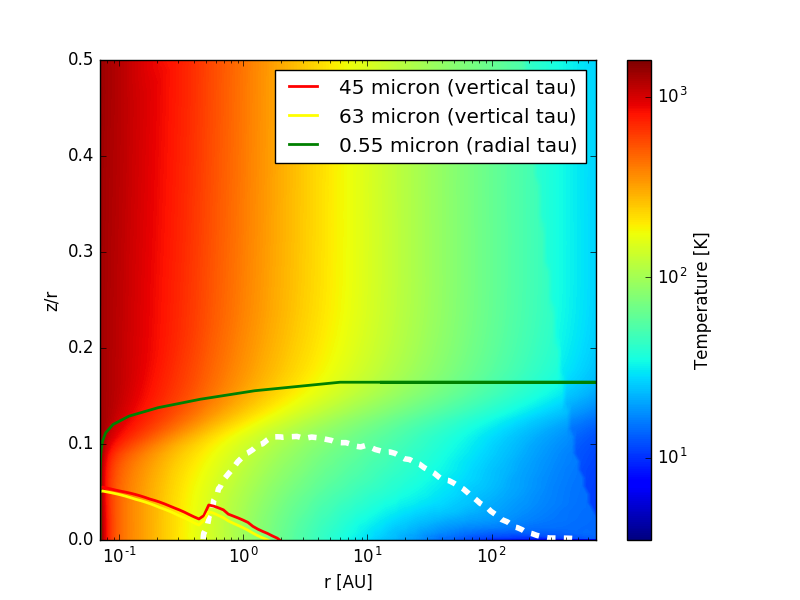} 
	\includegraphics[width=6cm]{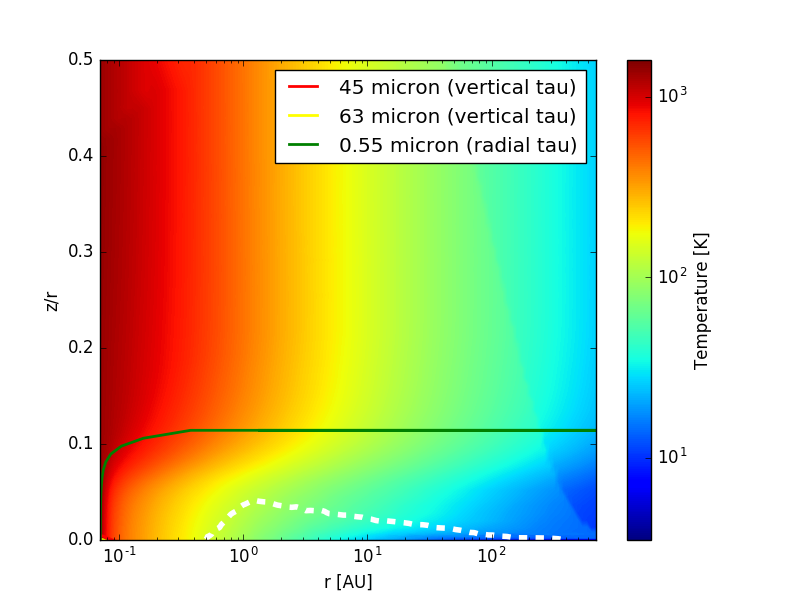} 
\caption{The 2D dust temperature distribution in the model series where the disk dust mass decreases from $10^{-3}$~M$_\odot$ to $10^{-7}$~M$_\odot$. The contour lines have the same meaning as in Fig.~\ref{fig:2Dtemp_referencemodel}.}
\label{fig:2Dtemp_Mdust}
\end{figure*}

The disks become increasingly transparent at mid-IR wavelength when the dust mass decreases from $10^{-3}$~M$_\odot$ to $10^{-7}$~M$_\odot$ (see red and yellow contour in Fig.~\ref{fig:2Dtemp_Mdust}). The full ice column density will be visible for dust masses below $10^{-4}$~M$_\odot$ beyond 100~au and for dust masses below $10^{-6}$~M$_\odot$ beyond 2~au. At the same time, the disk ice reservoir (white dashed contours) shrinks because the disk becomes warmer with decreasing dust mass and photodesorption can act at lower disk height due to an overall lower optical depth to UV photons. Especially in the lowest disk mass models, the change in opacity due to the presence of ice is now clearly visible in the optical depth contours.

\subsection{Ice fraction}
\label{App:sect:ice-fraction}

\begin{figure*}[htb]
	\includegraphics[width=9cm]{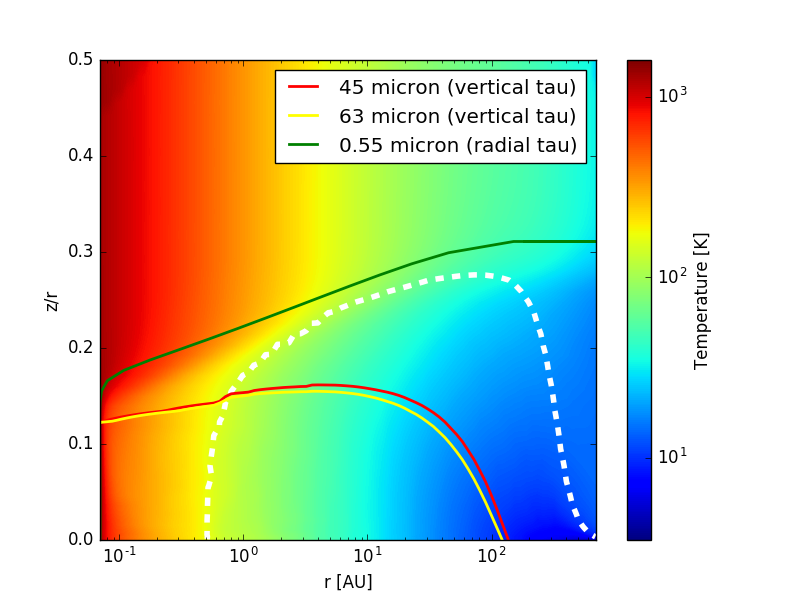} 
	\includegraphics[width=9cm]{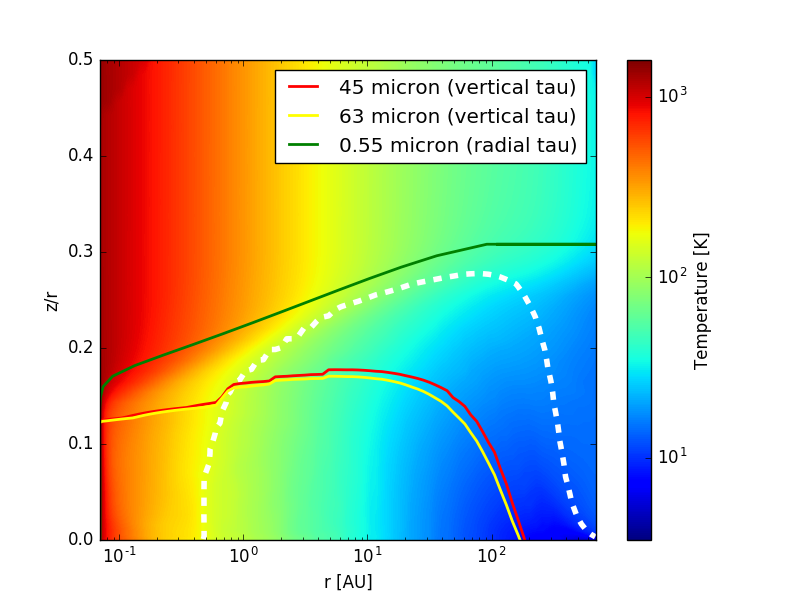} 
\caption{The 2D dust temperature distribution in the model series with different water ice fraction of 0.1 and 2. The contour lines have the same meaning as in Fig.~\ref{fig:2Dtemp_referencemodel}.}
\label{fig:2Dtemp_ice-fraction}
\end{figure*}

Figure~\ref{fig:2Dtemp_ice-fraction} clearly shows that the water ice fraction has a negligible impact on the disk thermal structure. It contributes very little to the overall grain energy balance. The explanation lies in the implementation, where the ice fraction determines an amount of ice to be added on top of the bare grain opacities. In that sense, changing the ice fraction does not change the underlying bare grain opacities. The grains become effectively thicker by growing more ice on the surface.

\subsection{Warmup versus cooldown scenario}
\label{App:sect:ice-scenarios}

\begin{figure*}[htb]
	\includegraphics[width=9cm]{2Dtemp_reference.png} 
	\includegraphics[width=9cm]{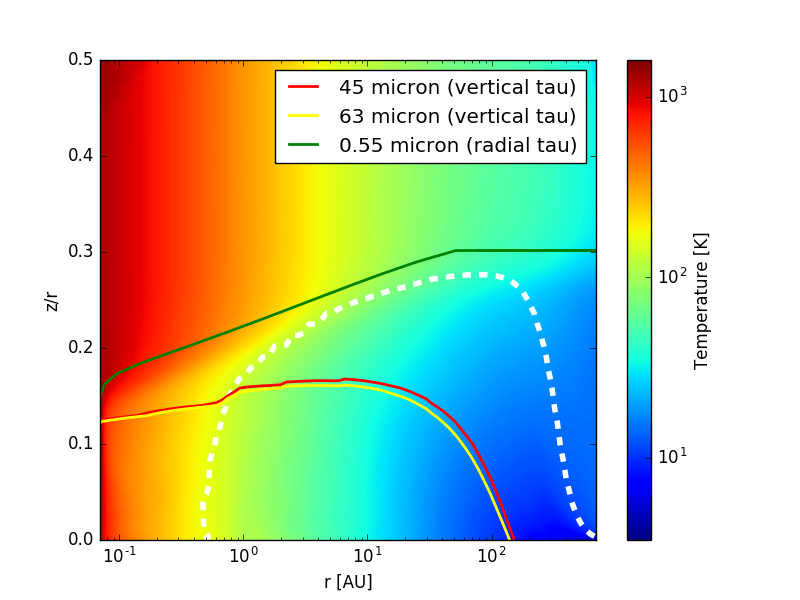} 
\caption{The 2D dust temperature distribution in the model series with different water ice `thermal histories'; for comparison, the first panel shows the reference disk model. The contour lines have the same meaning as in Fig.~\ref{fig:2Dtemp_referencemodel}.}
\label{fig:2Dtemp_ice-scenarios}
\end{figure*}

Figure~\ref{fig:2Dtemp_ice-scenarios} shows the difference between using fixed temperature 140~K crystalline ice opacities in the reference model and the full temperature dependent ice opacities from the cooldown series. While the temperature changes in the disk model are very small, there is a shift in the optical depth at $45~\mu$m to lower depth in the cooldown model. It is clear that the crystalline fixed 140~K ice opacity of the reference disk model systematically overestimates the ice opacity.

\end{appendix}

\bibliographystyle{aa}
\bibliography{references}

\end{document}